\documentclass[amsfonts, amsmath, amssymb, twocolumn, prd]{revtex4-1}
\usepackage{amsfonts}
\usepackage{graphicx}
\usepackage{dcolumn}
\usepackage{bm}

\usepackage{cleveref}
\usepackage{xcolor}

\date{\today}
\begin{document}



\title{Exact Renormalization Group, Entanglement Entropy, and Black Hole Entropy}



\author{Jo\~ao Lucas Miqueleto}\email{miqueleto.lucas@ufabc.edu.br}    
\affiliation{Center for Natural and Human Sciences, Federal University of ABC}

\author{Andr\'e G. S. Landulfo}\email{andre.landulfo@ufabc.edu.br}
\affiliation{Center for Natural and Human Sciences, Federal University of ABC}


\begin{abstract}
The study of black hole physics revealed a fundamental connection between thermodynamics, quantum mechanics, and gravity. Today, it is known that black holes are thermodynamical objects with well defined temperature and entropy. Although black hole radiance gives us the mechanism from which we can associate a well defined temperature to the black hole, the origin of its entropy remains a mystery. Here we investigate how the quantum fluctuations from the fields that render the black hole its temperature contribute to its entropy. By using the exact renormalization group equation for a self-interacting real scalar field in a spacetime possessing a bifurcate Killing horizon, we find the renormalization group flow of the total gravitational entropy. We show that throughout the flow one can split the quantum field contribution to the entropy into a part coming from the entanglement between field degrees of freedom inside and outside the horizon and a part due to the quantum corrections to the Wald entropy coming from the Noether charge. The renormalized black hole entropy is shown to be constant throughout the flow while the balance between the effective black hole entropy at low energies and the infra-red entanglement entropy changes. A similar conclusion is valid for the Wald entropy part of the total entropy.  Additionally, our calculations show that there  is  no mismatch  between  the  renormalization of the coupling constants coming from the effective action or the gravitational entropy, solving an apparent ``puzzle" that appeared to exist for interacting fields. 
\end{abstract}


\maketitle

\section{Introduction}\label{sec:I}

The discover of Hawking effect~\cite{hawking1975} unveiled  a deep connection between  black holes, quantum mechanics, and thermodynamics. In particular it made manifest that, when quantum effects are taken into account, black holes can be seen as real thermodynamic systems with temperature 
\begin{equation}
T_{\rm H}=\frac{\hbar c \kappa}{2\pi k_{\rm B}  G}
\label{Tbh}
\end{equation}
and entropy 
\begin{equation}
S^{\textrm{BH}}=\frac{k_{\rm B}c^{3} A_{\rm BH}}{4G\hbar},
\label{BHentropy}
\end{equation}
where $\kappa$ is the surface gravity of the black hole~\cite{wald84}, $A_{\rm BH}$ is the area of its event horizon,  and $k_{\rm B}$, $c$, $G$, and $\hbar$ are the Boltzmann constant, speed of light, Newton constant, and Planck constant, respectively. 

If on the one hand such discovery enriched our understanding of the nature of gravity and black holes, on the other one it raised new and intriguing questions concerning the quantum behavior of such objects. Among such puzzles, a special place is held for the question of what is the microscopic origin of the Bekenstein-Hawking entropy. We note that, although the appearance of $G,$ $c,$ and $\hbar$ in Eq.~(\ref{BHentropy}) suggests a quantum gravitational origin for the entropy, it does not gives us any hint of what degrees of freedom are being counted nor where they are located at. In fact, despite the numerous attempts to derive Eq.~(\ref{BHentropy}) from first principles using all kind of quantum gravity theories proposed over the years, the origin of the entropy of black holes remains elusive~\cite{Carlip_blackholethermodynamics_review, wall_blackholethermodynamics_reviewsurvey}. 
This has led many researchers to suggest that the Bekenstein-Hawking entropy may have a more familiar origin. 

As it is well known from the Hawking effect, the particle creation with thermal spectrum for all type of quantum fields is the key  ingredient responsible for associating a temperature to the black hole. Thus, one can expect that, if not all, at least part of its entropy has the same origin, i.e., it comes from the quantum fields present in the black hole spacetime. Considering the causal structure of such  spacetimes and the nature of Hawking radiation, it has been suggested that the black hole entropy could be explained by the entanglement between fields degrees of freedom inside and outside the black hole, the so-called entanglement entropy of black holes \cite{Solodukhin_review}. As the event horizon precludes an external observer of having access to all information about the state of the system, they will describe the field's state as a mixed state $\hat{\rho}_{\textrm{out}}$ which in turn will render the von Neumann entropy  
\begin{equation}
S^{\rm ent}\equiv-{\rm Tr}(\hat{\rho}_{\textrm{out}}\log\hat{\rho}_{\textrm{out}})
\label{SvN}
\end{equation}
a nonvanishing  value. Here, if  $|\psi\rangle$ is the total state of the field, $\hat{\rho}_{\textrm{out}}\equiv {\rm Tr}_{\rm in}|\psi\rangle \langle \psi|$ is the density matrix obtained after tracing out the field degrees of freedom inside the event horizon. As the total state  $|\psi\rangle$  is pure, Eq.~(\ref{SvN}) measures the entanglement between field degrees of freedom  inside and outside the black hole. 

When one compute Eq.~(\ref{SvN}) in the context of quantum field theories (QFTs) one obtains a divergence whose leading order term follows an area law. For example, the leading divergence for the entanglement entropy of a minimally-coupled free scalar field in a four-dimensional black hole spacetime is given by 
\begin{equation}
    S_{\textrm{div}}^{\rm ent}(\epsilon)=\frac{A_{\rm BH}}{48\pi\epsilon^{2}}+\dots,
    \label{eq:leading_order_divergence}
\end{equation}
where $\epsilon$ is an ultra-violet (UV) cutoff introduced to regularize the entropy. The next-to-leading order term is logarithmic divergent and it is represented by the dots in the above expression \cite{Solodukhin-conicalsingularity}. Such divergence is a common feature to almost all QFTs and its physical origin is the existence of arbitrarily short-distance correlations between the degrees of freedom inside and outside the black hole.

Within the scope of semiclassical gravity, the Bekenstein-Hawking entropy can be seen as a tree-level contribution \cite{gibbons_hawking_euclidean} while, as was pointed out by Callan and Wilczek in \cite{Callan_1994}, the entanglement entropy can be interpreted as the first quantum correction to the total black hole entropy.  However, all particle species and their interactions contribute to Eq.~(\ref{SvN}), which makes $S
^{\rm ent}$ dependent on the number of fundamental fields. The Bekenstein-Hawking entropy, in turn, depends only on the physical value of Newton constant $G$. This is the so-called ``species problem" and it can be solved in a natural manner if the renormalization of the divergences appearing in the entanglement entropy, $S
^{\rm ent},$ match the renormalization of Newton constant $G$. This, in fact, turns out to be the case for minimally-coupled free scalar and spinor fields~\cite{Solodukhin_review}. For gauge fields~\cite{Solodukhin_review, kabat, solodukhin12, donnelly_wall} and non-minimally coupled scalar field~\cite{Solodukhin_oneloop}, there appears to be a discrepancy due to the appearance of an extra ``contact" term in the divergence of $1/G$ when compared to the divergence of the entanglement entropy.  
For the scalar field case, which we will be more interested in the present paper, if there is a non-minimal coupling $\xi R\phi^{2}$ between the quantum field $\phi$ and the Ricci scalar curvature $R$, the leading-order divergence  for the entropy is given by \cite{Solodukhin_oneloop}
\begin{equation}
    S_{\textrm{div}}^{\xi}(\epsilon)=\frac{A_{\rm BH}}{8\pi\epsilon^{2}}\left(\frac{1}{6}-\xi\right)+\dots,
    \label{eq:leading_order_term_entropy_nonminimalcoupling}   \end{equation}
where $\xi\in\mathbb{R}$. This extra contact term appearing due to the non-minimal coupling does not have a statistical interpretation in terms of the von Neumann entropy. In fact, depending on the value of $\xi$, Eq.~(\ref{eq:leading_order_term_entropy_nonminimalcoupling}) can become negative. Nevertheless, as pointed out by Donnelly and Wall~\cite{donnelly_wall}, due to the presence of such non-minimal coupling, the classical gravitational entropy must be seen as 
\begin{equation}
 S^{\textrm{grav}}=\frac{A_{\rm BH}}{4G} + S^{\rm wald},
    \label{eq:bare_gravitational_entropy_interaction_introduction}
\end{equation}
where 
\begin{equation}
S^{\rm wald}\equiv -2\pi\xi\int_{\Sigma}\sqrt{\gamma}d^{2}\textbf{x}\varphi^{2}(\textbf{x})
\label{S_wald_intro}
\end{equation}
is the Wald entropy~\cite{waldentropy}, with $\Sigma$ being the event horizon bifurcation surface and $\gamma$ is the determinant of the induced metric on $\Sigma$ written in coordinates ${\bf x}$ covering the bifurcation surface. As a result, they have shown that the contact term is in fact a quantum correction to Wald's entropy while the usual divergent area term gives the leading entanglement entropy correction to the black hole entropy. 

Unfortunately,  this nice picture appears to break down when one considers a self-interacting and non-minimally coupled scalar field.  In \cite{Solodukhin_nonrelativistic} it is argued that an interaction does not affect the leading order UV divergence in the entropy, which remains equal to the one given in Eq.~(\ref{eq:leading_order_divergence}). The effect of a $\phi^4$ self-interaction appears only in the sub-leading logarithm term as
\begin{eqnarray}
S_{\textrm{div}}^{\rm ent}(\epsilon)=\frac{A_{\rm BH}}{48\pi\epsilon^{2}}+\frac{\lambda}{24\pi}\log\epsilon\int_{\Sigma}\sqrt{\gamma} d^{2}\textbf{x}\varphi^{2}(\textbf{x}),
\label{eq:entropy_div_interaction_introduction}
\end{eqnarray}
where $\varphi=\langle \phi\rangle $ is the classical field and $\lambda$ is the self-interaction coupling constant. 
This leads again to a mismatch between the renormalization of the coupling constants (now $\xi$) and the divergence of the entropy~\cite{Solodukhin_review,Solodukhin_nonrelativistic}. The renormalization of $\xi$, which is known to be given by~\cite{NP82}  
\begin{eqnarray}
\xi_{\textrm{ren}}=\xi-\frac{\lambda}{8\pi^{2}}\left(\frac{1}{6}-\xi\right)\log\epsilon,
\label{eq:non-minimal_coupling_renormalized_interactingfieldtheory}
\end{eqnarray}
will not render the total entropy $S^{\textrm{grav}}+S^{\rm ent}_{\textrm{div}}(\epsilon)$ finite. This poses a serious difficulty in trying to explain the black hole entropy by means of the entanglement entropy.

In the present paper, we will investigate the possibility of interpreting (at least part) of the Bekenstein-Hawking entropy as coming from the entanglement entropy by looking at the problem through a (somewhat) different angle. Instead of dealing with UV divergences throughout the calculations, it would be more interesting if we could study the entropy using only manifestly finite quantities. In order to do so, we will make use of the exact renormalization group (ERG)
~\cite{Wetterich_1993,reuter_saueressig_2019}, which employs the introduction of an arbitrary variable infra-red (IR) energy cutoff $k^2$ in the theory that will allow us to divide the field modes in terms of high-energy and low-energy modes. When high-energy modes are integrated out, we will obtain an effective field theory for the modes below $k^{2}$ described by a $k$-dependent effective action, also known as effective average action (EAA). This EAA interpolates between the full quantum effective action, when $k\to 0$, and the bare action, when $k\to\infty$. The advantage to use the EAA is that it satisfies an exact renormalization group equation (ERGE) from which one can obtain the flow equations of the couplings directly. In addition, by using this formalism, it will be possible to analyze how the balance between the effective gravitational entropy and its quantum corrections changes as one varies $k$ and how this influences the total gravitational entropy.

This approach to study the entanglement entropy of black holes was first used by Jacobson and Satz in~\cite{jacobson_satz}.  There, they have studied the (on-shell) flow of the gravitational entropy for minimally-coupled free scalar field theory in an Euclidean Schwarzschild black hole. Here, in turn, we take a step further to address the (off-shell) case of a non-minimally coupled self-interacting scalar field theory in spacetimes possessing a bifurcate Killing horizon. By analyzing the ERGE for the total gravitational entropy, we will show that {\bf (1)} there will be a well-behaved balance between the effective gravitational entropy at a scale $k$ and its quantum corrections below $k$ and {\bf (2)} the renormalization group (RG) flow of the coupling constants coming from the effective action matches exactly the RG flow of the coupling constants coming from the entropy, solving the apparent problem that it was believed to exist in the interacting case. As a result, even for a self-interacting and non-minimally coupled scalar field, the entanglement entropy is a viable description of the origin of the Bekenstein-Hawking entropy.

The paper is organized as follows. In Sec.~\ref{sec:II} we review the conical method used to compute both the classical  gravitational entropy and the entropy associated with the quantum fluctuations of the field. In Sec.~\ref{sec:III} we will derive the  ERG flow for the gravitational entropy. Then, in Sec.~\ref{sec:IV}, we take the 1-loop approximation of the ERG flow to show how the total entropy is divided into an effective entropy at a scale $k$ and a quantum contribution coming from modes below $k$ and how the balance between such contributions changes as $k$ is varied. In addition, we solve the apparent mismatch between the renormalized non-minimal coupling and entropy renormalization and show that, once the QFT is renormalized, the entropy comes out automatically finite. We summarize our conclusions in Sec.~\ref{sec:V}. We adopt metric signature $(-,+,+,+)$ and natural units, $\hbar=c=k_{\rm B}=1$, unless stated otherwise.

\section{The Conical Method and the Entanglement Entropy}
\label{sec:II}

\begin{figure}
\begin{center}
\includegraphics[scale=0.35]{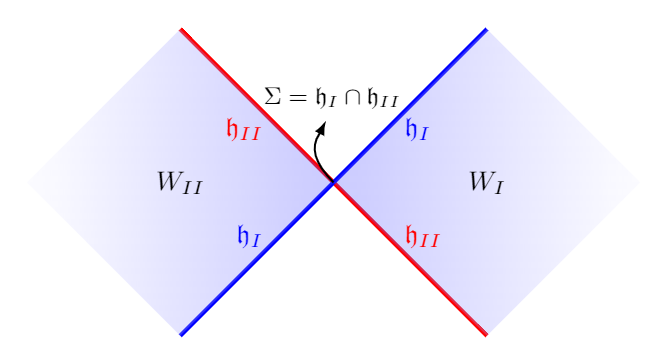}
\end{center}
\caption{Bifurcated Killing horizon  $\mathfrak{h}=\mathfrak{h}_I\cup \mathfrak{h}_{II}$ with bifurcation surface $\Sigma$ on $(\mathcal{M}_{L}, g_{ab}^{L})$. The regions $W_I$ and $W_{II}$ are defined as the exterior and interior region to the horizon, respectively.}
\label{fig1}
\end{figure}

Let us consider a globally hyperbolic spacetime $(\mathcal{M}_{L}, g_{ab}^{L})$, where $\mathcal{M}_{L}$ is a four-dimensional manifold  equipped with a Lorentzian metric $g_{ab}^{L}$, containing a bifurcate Killing horizon $\mathfrak{h}\equiv \mathfrak{h}_I\cup \mathfrak{h}_{II} $ with bifurcation surface $\Sigma=\mathfrak{h}_I\cap \mathfrak{h}_{II}$ \cite{wald1994quantum}.  The horizon $\mathfrak{h}$ divides the spacetime into four wedges with wedge $I$,
\begin{equation}
W_I\equiv I^-\left(\mathfrak{h}_I\right)\cap I^+\left(\mathfrak{h}_{II}\right),
\label{WI}
\end{equation}  and wedge $II$, 
\begin{equation}
W_{II}\equiv I^+\left(\mathfrak{h}_I\right)\cap I^-\left(\mathfrak{h}_{II}\right),
\label{WII}    
\end{equation}
being defined as the causally disconnected regions exterior and interior to the horizon, respectively (see Fig.~\ref{fig1}). Here, $I^{+}(S)$ and $I^{-}(S)$ indicate the chronological future and past of a region $S\subset\mathcal{M}_L,$ respectivelly~\cite{wald84}.

In such spacetimes, any gravity theory described by a diffeomorphism invariant Lagrangian 
\begin{equation}
    L\equiv L(g^L_{ab}, R_{abcd}, \nabla_aR_{bcde},\cdots; \Psi, \nabla_a\Psi,\cdots),
    \label{diffInvL}
\end{equation}
where $\nabla_a$ is the torsion-free covariant derivative compatible with $g_{ab}^{L}$, $R_{abcd}$ is the curvature associated with $\nabla_a$, and $\Psi$ represents any matter fields present, satisfies the first law of black hole mechanics for arbitrary perturbations to nearby stationary black hole solutions~\cite{waldentropy}. The entropy in this case will be given by 

\begin{equation}
    S^{\rm grav}=-2\pi \int_\Sigma\sqrt{\gamma}d^2{\bf x}\frac{\delta L}{\delta R^{abcd}}\epsilon^{ab}\epsilon^{cd},
    \label{generalWald}
\end{equation}
where $\epsilon^{ab}$ is the bi-normal to $\Sigma$ satisfying $\epsilon^{ab}\epsilon_{ab}=-2$, ${\bf x}$ are coordinates on $\Sigma$, and $\gamma$ is the determinant  of the induced metric $\gamma_{ab}$ on $\Sigma$. Moreover, if any matter field is quantized its quantum fluctuations would contribute to the total entropy.  Therefore, if $\phi$ is a quantum field prepared in a pure state $|\psi\rangle$, observers restricted to region $W_I$ will describe it as 
$$\hat{\rho}_{\textrm{out}}\equiv-{\rm Tr}_{W_{II}}|\psi\rangle\langle \psi|,$$
which is usually a mixed state with von Neumann entropy
\begin{equation}
    S_{\hat{\rho}}\equiv-{\rm Tr}\left(\hat{\rho}_{\textrm{out}}\log\hat{\rho}_{\textrm{out}}\right).
    \label{eq:entanglement_entropy_von_neumann}
\end{equation}

Although Eq.~(\ref{eq:entanglement_entropy_von_neumann}) has a clear physical meaning--it quantifies the entanglement between field degrees of freedom in regions $W_I$ and $W_{II}$--it is not adequate to perform calculations when one is dealing with QFTs. Next, we will cast it in a form more suitable for dealing with states in QFTs. Additionally, this will also give us an alternative way to compute the gravitational entropy~(\ref{generalWald}). Such unified formalism for computing Eqs.~(\ref{generalWald}) and~(\ref{eq:entanglement_entropy_von_neumann}) will be very useful in describing the ERG flow of the total entropy. 

For the sake of our calculations, it will be convenient to consider the analytical continuation of $(\mathcal{M}_L, g^L_{ab})$ to an Euclidean spacetime $(\mathcal{M}_E, g^E_{ab})$.  When this is done, the Lorentzian Killing field generating $\mathfrak{h}$ becomes the generator of rotations around $\Sigma$. From now on, all our analysis will be done in the Euclidean section of the spacetime and we will omit the index ``E'' of Euclidean quantities unless stated otherwise.

We will rewrite Eq.~(\ref{eq:entanglement_entropy_von_neumann}) in a more suitable form by means of the so-called conical method (or replica trick)~\cite{Callan_1994, Calabrese_2004,Casini_2009}. For this purpose, let us first note that we can write the von Neumann entropy as 
\begin{equation}
S_{\hat{\rho}} = \lim_{n\rightarrow 1}S_n(\hat{\rho})
    \label{vNR1}
\end{equation}
where 
\begin{equation}
    S_{n}(\hat{\rho})\equiv \frac{1}{1-n}\log{\rm Tr}\hat{\rho}^{n},
    \label{eq:renyi_entropy}
\end{equation}
with $n\in\mathbb{N}$ and $n>1$, is the R\'enyi entropy of a state $\hat{\rho}$~\cite{renyi1961,renyi1965}. From Eq.~(\ref{eq:renyi_entropy}) it easy to see that we can cast the entropy~(\ref{vNR1}) in the form 
\begin{equation}
    S_{\hat{\rho}}=-\frac{\partial}{\partial n}\log{\rm Tr}\hat{\rho}^{n}\bigg|_{n=1}
    \label{eq:intermediate_von_neumann_entropy}
\end{equation}
which, by means of an analytical continuation to complex values, $n\to\alpha$, with $\alpha\in\mathbb{C}$ and $\Re(\alpha)>1$, can be written as~\cite{Solodukhin_review,Nesterov_solodukhin}
\begin{equation}
    S_{\hat{\rho}}=-\left(\alpha\frac{\partial}{\partial\alpha}-1\right)\log{\rm Tr}\hat{\rho}^{\alpha}\bigg|_{\alpha=1}.
    \label{eq:intermediate_von_Neumann_entropy_alpha}
\end{equation}
Now, for the case we are considering, we can write ${\rm Tr\hat{\rho}^\alpha}$ in terms of an $\alpha$-dependent effective action $\Gamma(\alpha)$ satisfying~\cite{Solodukhin_review} 
\begin{equation}
   e^{-\Gamma(\alpha)}={\rm Tr}\hat{\rho}^{\alpha},
    \label{eq:generating_functional_alphaparameter}
\end{equation}
with which we can cast Eq.~(\ref{eq:intermediate_von_Neumann_entropy_alpha}) as 
\begin{equation}
    S_{\hat{\rho}}=\left(\alpha\frac{\partial}{\partial\alpha}-1\right)\Gamma(\alpha)\bigg|_{\alpha=1}.
    \label{eq:final_entropy_formula_depending_alpha_effectiveaction}
\end{equation}
Note that, in order to write the entropy as in Eq.~(\ref{eq:final_entropy_formula_depending_alpha_effectiveaction}), we have considered the $\alpha$-fold covering $\left(\mathcal{M}_\alpha, g_{ab}(\alpha)\right)$ of $(\mathcal{M}, g_{ab})$. The manifold  $\mathcal{M}_\alpha$ has a conical singularity at $\Sigma$ with angle deficit $\delta=2\pi(1-\alpha)$ which implies that the scalar curvature picks up a $\delta_{\Sigma}$ singularity~\cite{Fursaev_1995}
\begin{equation}
    R\rightarrow \bar{R} + 4\pi (1-\alpha)\delta_{\Sigma},
    \label{eq:curvature_scalar:conical_geometry}
\end{equation}
where $\delta_{\Sigma}$ is the Dirac delta distribution on the bifurcation surface $\Sigma$ and $\bar{R}$ is the regular part of the scalar curvature, i.e., the value of $R$ away from $\Sigma$ (which is the value of $R$ in $\mathcal{M}$).  

Now, as it is shown in~\cite{IW95} (see also~\cite{Solodukhin_review} and references therein), when one considers the analytic continuation of the Lagrangian~(\ref{diffInvL}) to $\left(\mathcal{M}_\alpha, g_{ab}(\alpha)\right)$, it is possible to write the classical gravitational entropy~(\ref{generalWald}) as 
\begin{equation}
S^{\rm grav}=\left.\left(\alpha\frac{\partial}{\partial\alpha}-1\right)\Gamma_\Lambda\left[g_{ab}(\alpha), \Psi\right]\right|_{\alpha=1}
    \label{SgravConical}
\end{equation}
where 
\begin{eqnarray}
&&\Gamma_\Lambda\left[g_{ab}(\alpha), \Psi\right]\equiv \int_{\mathcal{M}_\alpha}\sqrt{g(\alpha)} d^4x\times \nonumber \\
&&  L(g_{ab}(\alpha), R_{abcd}(\alpha), \nabla^\alpha_aR_{bcde}(\alpha),\cdots; \Psi, \nabla^\alpha_a\Psi,\cdots),\nonumber \\
    \label{classAction}
\end{eqnarray}
with $\nabla_a^\alpha$ being the torsion-free covariant derivative compatible with $g_{ab}(\alpha)$~\footnote{It interesting to note that Eq.~(\ref{generalWald}) is derived on-shell while the conical method is an off-shell procedure which is valid in any spacetime with a bifurcate Killing horizon. For the relation between the on- and off-shell approaches, see for instance, Ref.~\cite{Solodukhin_review}}.

As a result, the conical method enables one to write in an unified way the classical gravitational entropy, given in Eq.~(\ref{SgravConical}), and its quantum corrections, given in Eq.~(\ref{eq:final_entropy_formula_depending_alpha_effectiveaction}) (which amounts to the entanglement entropy in the minimally coupled case). This form will be particularly useful in the next section when we derive the ERG flow of the total entropy. 

\section{The Renormalization Group Flow For the Gravitational Entropy}\label{sec:III}

Let us consider a non-minimally coupled and self-interacting real scalar field $\phi$ propagating in $\left(\mathcal{M},g_{ab}\right)$. Here, we will use the ERGE as a tool to perform the calculations of the entropy flow. The main object of this equation is the EAA,  $\Gamma_{k}[g_{ab},\varphi],$ which depends on an  energy scale $k^2$, where $\varphi\equiv \langle \phi\rangle$ is the vacuum expectation value of the field $\phi$. The EAA satisfies the ERGE~\cite{Wetterich_1993}
\begin{equation}
    k\partial_k \Gamma_k=\frac{1}{2}{\rm sTr}\left[\left(\Gamma^{(2)}_k + R_k(\nabla^2)\right)^{-1}k\partial_k R_k(\nabla^2)\right],
    \label{Eq:ERGE}
\end{equation}
where $\nabla^2\equiv -\nabla^a\nabla_a$,
$$\Gamma^{(2)}_k\equiv \frac{\delta^2 \Gamma_k}{\delta \varphi \delta \varphi},$$
and
\begin{equation}
R_k(p^2)=k^2\left(1- \frac{p^2}{k^2}\right)\theta\left(1-\frac{p^2}{k^2}\right )
    \label{eq:litim_cut-off}
\end{equation} 
is a cut-off function, with $\theta(x)$ being the Heaviside function. Equation~(\ref{Eq:ERGE}) contains all the information about the QFT. The solution of such equation is an element of the so-called theory space~\cite{reuter_saueressig_2019}, defined as the  space of all possible functionals of the field respecting the underlying symmetries. However, it is not possible to solve Eq.~(\ref{Eq:ERGE}) exactly and thus some approximation method must be employed. One of the main advantages of using the ERGE is the possibility of adopting non-perturbative methods in a consistent way. In our case, we will project our EAA onto a  submanifold of the theory space~\footnote{It is important to note that restricting to such a submanifold in the theory space will induce a restriction of the entropy flow to a submanifold in the space of entropies.}, i.e., we will choose a truncation in the series expansion of $\Gamma_k$ in which it takes the form
 
\begin{eqnarray}
\Gamma_{k}[g_{ab},\varphi]=\Gamma_{k}^{\textrm{EH}}[g_{ab}]+\Gamma_{k}^{\phi}[g_{ab},\varphi]+\Gamma_{k}^{\textrm{int}}[g_{ab},\varphi],
\label{eq:definition_EAA_general}
\end{eqnarray}
where 
\begin{eqnarray}
\Gamma_{k}^{\textrm{EH}}[g_{ab}]=-\frac{1}{16\pi G_{k}}\int_{\mathcal{M}}\sqrt{g} d^{4}x \left(R-2\Lambda_{k}\right)
\label{eq:Einstein-Hilbert action}
\end{eqnarray}
is the Einstein-Hilbert action, 
\begin{equation}
\!\!\!\Gamma^{\phi}_{k}[g_{ab},\varphi]=\frac{1}{2}\int_{\mathcal{M}}\sqrt{g} d^{4}x \;\varphi\left(-\nabla^{a}\nabla_{a}+\xi_{k}R+m^{2}_{k}\right)\varphi
\label{eq:free_scalar_field_action}
\end{equation}
is the free part of the scalar field effective action, and
\begin{eqnarray}
\Gamma^{\textrm{int}}_{k}[g_{ab},\varphi]=\frac{\lambda_{k}}{4!}\int_{\mathcal{M}}\sqrt{g} d^{4}x \; \varphi^{4}
\label{eq:interaction_part_scalar_field}
\end{eqnarray}
is the self-interaction effective action for $\phi$. Here $G_{k}$, $\Lambda_{k}$, $m_{k}$, $\xi_{k},$ and $\lambda_{k}$ are the Newton constant, cosmological constant, field mass, non-minimal coupling of the field with gravity, and the self-interaction coupling constant, respectively. Note that all couplings depend on the scale $k$. Here, we are splitting the scalar quantum field as $$\phi=\varphi+\delta\phi,$$ where $\varphi=\langle \phi\rangle $ is the background field configuration and $\delta\phi$ is the (not necessarily small) perturbation around $\varphi$. The field state is chosen such that the perturbation field $\delta \phi$ is in the Hartle-Hawking state $|0_{\textrm{HH}}\rangle $~\cite{harte_hawking_original}, which is the unique renormalizable state invariant under the symmetries generating $\mathfrak{h}$~\cite{jacobson_hartle_hawking_vacua, WK91}. In what follows, the background metric $g_{ab}$ will be kept fixed while $\delta \phi$ is quantized on it.

Let us now use Eqs.~(\ref{SgravConical}) and~(\ref{Eq:ERGE}) to compute the flow of the gravitational entropy at a scale $k$
\begin{equation}
    S^{\rm grav}_k\equiv \left. D_{\alpha} \Gamma_k[g_{ab}(\alpha), \varphi]\right|_{\alpha=1},\label{eq:gravitational_entropy_general_formula_conicalspace}
\end{equation}
where  
\begin{equation}
    D_{\alpha}\equiv (\alpha \partial_\alpha -1)
    \label{Dalpha}
\end{equation} 
and $\Gamma_{k}[g_{ab}(\alpha),\varphi]$ is the EAA~(\ref{eq:definition_EAA_general}) evaluated on the conical space $(\mathcal{M}_{\alpha},g_{ab}(\alpha))$. By using Eqs.~(\ref{eq:curvature_scalar:conical_geometry}) and~(\ref{eq:Einstein-Hilbert action})-(\ref{eq:interaction_part_scalar_field})  in Eq.~(\ref{eq:definition_EAA_general}) and making use  Eq.~(\ref{eq:gravitational_entropy_general_formula_conicalspace}), we find that the gravitational entropy is given by

\begin{equation}
    S^{\rm grav}_k =\frac{A_{\Sigma}}{4G_k} -2\pi \xi_k\int_{\Sigma}\sqrt{\gamma}d^2{\bf x} \varphi^2({\bf x}),
    \label{Sgrav}
\end{equation}
where $A_\Sigma$ is the area of $\Sigma.$ We can see from the above equation that the first term corresponds to the black hole contribution to the entropy while the second term is the Wald entropy coming from the non-minimal coupling of the field with gravity. Therefore, 
\begin{equation}
    S_{k}^{\textrm{grav}}=S_{k}^{\textrm{BH}}+S_{k}^{\textrm{wald}}
    \label{eq:Sgrav=Sbh+Swald}
\end{equation}
with
\begin{eqnarray}
S^{\rm BH}_k & \equiv &\frac{A_{\Sigma}}{4G_k}, 
\label{BH}\\
S^{\rm wald}_k&\equiv&-2\pi \xi_k\int_{\Sigma}\sqrt{\gamma}d^2{\bf x}~\varphi^2({\bf x}).
\label{SWald}
\end{eqnarray}

The RG flow of the gravitational entropy will be obtained by acting on both sides of the Eq.~(\ref{eq:gravitational_entropy_general_formula_conicalspace}) with $k\partial_{k}$ and then using  Eq.~(\ref{Eq:ERGE}),  which will appear on its right-hand side. In order to do so, let us first analyze the form that Eq.~(\ref{Eq:ERGE}) takes on $\left(\mathcal{M}_\alpha,g_{ab}(\alpha)\right)$. By using that 
\begin{equation}
    \Gamma^{(2)}_k[g_{ab}(\alpha), \varphi]= \nabla^2 + m^2_k+ \frac{\lambda_k\varphi^2}{2} + \xi_k \bar{R} + 4\pi (1-\alpha)\xi_k\delta_{\Sigma}
\end{equation}
and defining 
$$P_{k}\left(\nabla^2\right)\equiv \nabla^2 + R_k(\nabla^2)$$ and $$q_k\equiv m^2_k + \lambda_k\varphi^2/2,$$ we can cast Eq.~(\ref{Eq:ERGE}) on $\mathcal{M}_\alpha$ as 

\begin{eqnarray}
     &&k\partial_k \Gamma_k=\frac{1}{2}{\rm sTr}_{\mathcal{M}_\alpha}\left[\frac{k\partial_k R_k(\nabla^2)}{P_{k}\left(\nabla^2\right) + q_k + \xi_k \bar{R}}\right] \nonumber \\
     &&-2\pi \xi_k(1-\alpha)\:{\rm sTr}_\Sigma\left[\frac{k\partial_k R_k(\nabla^2)}{\left(P_{k}\left(\nabla^2\right) + q_k + \xi_k \bar{R}\right)^{2} }\right]\nonumber \\ &&+\mathcal{O}\left((1-\alpha)^2\right), 
     \label{eq:RHS_ERGE_conicalspace}
\end{eqnarray}
where ${\rm sTr}_{\mathcal{M}_\alpha}$ and ${\rm sTr}_\Sigma$ indicate that the functional traces are taken on  $\mathcal{M}_\alpha$ and $\Sigma$, respectively. Now, let us  apply $D_{\alpha}|_{\alpha=1}$ to Eq.~(\ref{eq:RHS_ERGE_conicalspace}) and use Eq.~(\ref{eq:gravitational_entropy_general_formula_conicalspace}) to write 
\begin{align}\nonumber
    k\partial_{k}S_{k}^{\textrm{grav}}&=\left.\frac{1}{2}D_{\alpha}\textrm{sTr}_{\mathcal{M}_{\alpha}}\left[\frac{k\partial_{k}R_{k}\left(\nabla^2\right)}{P_{k}\left(\nabla^2\right)+q_{k}+\xi_{k}\bar{R}}\right]\right|_{\alpha=1}&\\ 
    &+2\pi\xi_{k}\:\textrm{sTr}_{\Sigma}\left[\frac{k\partial_{k}R_{k}\left(\nabla^2\right)}{\left(P_{k}\left(\nabla^2\right)+q_{k}+\xi_{k}\bar{R}\right)^2}\right],&
     \label{eq:RenormalizationGroupFlow_Gravitational_Entropy}
\end{align}
which is the desired ERGE flow of the gravitational entropy. Note that, in order to be consistent with our approximation, we  will need to project Eq.~(\ref{eq:RenormalizationGroupFlow_Gravitational_Entropy}) onto our (induced) entropy truncation subspace. To do so, let us expand the right-hand side of the ERGE~(\ref{eq:RenormalizationGroupFlow_Gravitational_Entropy}) in powers of $\varphi^{2}$ and $\bar{R}$ as
\begin{eqnarray}
   &&k\partial_k S^{\rm{grav}}_k=\frac{1}{2}D_{\alpha}\left.{\rm sTr}_{\mathcal{M}_\alpha}\left[\frac{k\partial_k R_k\left(\nabla^2\right)}{P_{k}\left(\nabla^2\right)+ m_k^2}\right]\right|_{\alpha=1} \nonumber \\
    &&-\frac{\lambda_k}{4}  D_{\alpha}\left.{\rm sTr}_{\mathcal{M}_\alpha}\left[\frac{\varphi^2 k\partial_k R_k\left(\nabla^2\right)}{\left(P_{k}\left(\nabla^2\right) + m_k^2\right)^{2}}\right]\right|_{\alpha=1} \nonumber \\
     &&+2\pi \xi_k\left.{\rm sTr}_\Sigma\left[\frac{k\partial_k R_k\left(\nabla^2\right)}{\left(P_{k}\left(\nabla^2\right) + m_k^2 \right)^{2}}\right]\right.\nonumber \\
 &&-2\pi \xi_k\lambda_k\left.{\rm sTr}_\Sigma\left[ \frac{\varphi^2 k\partial_k R_k\left(\nabla^2\right)}{\left(P_{k}\left(\nabla^2\right) + m_k^2 \right)^{3}}\right]\right.\nonumber \\
 &&+ \mathcal{F}_{k}\left(m^2_k,\lambda_k\varphi^2,\xi_k\bar{R}\right) +\mathcal{O}\left(\lambda_k^2\varphi^4,\xi_k^2\bar{R}^2\right), 
\label{Eq:ERG_Flow_SGrav_expanded}
\end{eqnarray}
where  
\begin{widetext}
\begin{eqnarray}\nonumber
  \mathcal{F}_{k}\left(m^2_k,\lambda_k\varphi^2,\xi_k\bar{R}\right)&&\equiv-\frac{\xi_k}{2}D_{\alpha}\left.{\rm sTr}_{\mathcal{M}_\alpha}\left[\frac{\bar{R}k\partial_k R_k\left(\nabla^2\right)}{\left(P_{k}\left(\nabla^2\right)+ m_k^2\right)^{2}}\right]\right|_{\alpha=1} +\frac{\lambda_k\xi_k}{2}  D_{\alpha}\left.{\rm sTr}_{\mathcal{M}_\alpha}\left[\frac{\varphi^2\bar{R}k\partial_k R_k\left(\nabla^2\right)}{\left(P_{k}\left(\nabla^2\right)+ m_k
 ^2\right)^{3}}\right]\right|_{\alpha=1}\nonumber \\ &&-4\pi \xi_k^2 \left.{\rm sTr}_\Sigma\left[\frac{\bar{R}k\partial_k R_k\left(\nabla^2\right)}{\left(P_{k}\left(\nabla^2\right) + m_k^2 \right)^{3}}\right]\right. +6\pi \xi_k^2\lambda_k\left.{\rm sTr}_\Sigma\left[\frac{\varphi^2\bar{R}k\partial_k R_k\left(\nabla^2\right)}{\left(P_{k}\left(\nabla^2\right) + m_k^2 \right)^{4}}\right]\right. .
\label{eq:function_Fk}
\end{eqnarray}

By using the Heat Kernel expansion of $\nabla^2 $, the supertraces in Eq.~(\ref{Eq:ERG_Flow_SGrav_expanded}) can be computed by using the following expansions~\cite{reuter_saueressig_2019,Percacci_Book}: 

\begin{eqnarray}
   \textrm{sTr}_{\mathcal{M}_{\alpha}}\left[f\left(P_{k}(\nabla^{2})+m^{2}_{k}\right)^{-l}k\partial_{k}R_{k}(\nabla^{2})\right]&=&\frac{1}{(4\pi)^{2}}\sum_{k=0}^{\infty}Q_{2-k}\left[W_{l}(\nabla^{2})\right]A_{k}^{\mathcal{M}_{\alpha}}\left[f,\nabla^{2}\right]\label{eq:sTr_1},\\ 
   \textrm{sTr}_{\Sigma}\left[f\left(P_{k}(\nabla^{2})+m^{2}_{k}\right)^{-l}k\partial_{k}R_{k}(\nabla^{2})\right]&=&\frac{1}{(4\pi)^{2}}\sum_{k=0}^{\infty}Q_{2-k}\left[W_{l}(\nabla^{2})\right]A_{k}\left[f\delta_{\Sigma},\nabla^{2}\right]\label{eq:sTr_2},
\end{eqnarray}
\end{widetext}
where $l\in\mathbb{N}$, $f$ is a scalar function, and $W$ is given by 

\begin{eqnarray}
   W(z)&=&\frac{k\partial_{k}R_{k}(z)}{\left(P_{k}(z)+q_{k}\right)^{l}}=\frac{2k^{2(1-l)}\theta(1-z)}{\left(1+k^{-2}m_{k}^{2}\right)^{l}},
   \label{eq:W_function}
\end{eqnarray}
with $R_{k}(z)$ being given in Eq.~(\ref{eq:litim_cut-off}) and $z\equiv p^{2}/k^{2}$ being a dimensionless variable. The  Q-functionals are defined as~\cite{reuter_saueressig_2019,Percacci_Book} 
\begin{equation}
   Q_{n}[W(\Delta)]=\frac{1}{\Gamma(n)}\int_{0}^{\infty}dz z^{n-1}W(z),
   \label{eq:Q_functionals_general}
\end{equation}
where $\Gamma(n)$ is the gamma function, and with the choice~(\ref{eq:W_function}) of cut-off function $R_k$ they take the form
\begin{equation}
Q_n\left[W_l(\nabla^2)\right]=\left\{
\begin{array}{cc} 
2k^{2(n-l+1)}/n!\left(1+k^{-2}m^2_k\right)^{l},  & n\geq 0  \\
\:\:\:\:\:\:\:\:\:\:\:0\:\:\:\:\:\:\:\:\:\:\:\:\:,  & n<0 
\end{array}
\right..
\label{eq:Q-Functional_ourcase}
\end{equation}
The coefficients $A_{k}^{\mathcal{M}_{\alpha}}[f,\nabla^{2}]$ and $A_{k}[f\delta_{\Sigma},\nabla^{2}]$ are given by~\cite{Solodukhin_review} 
\begin{widetext}
\begin{eqnarray}
A^{\mathcal{M}_\alpha}_0\left[f,\nabla^2\right]&=&\int_{\mathcal{M}_\alpha}\sqrt{g(\alpha)}d^4x f, \label{eq:AM_1}\\ 
A^{\mathcal{M}_\alpha}_1\left[f,\nabla^2\right]&=& \frac{1}{6}\int_{\mathcal{M_\alpha}}\sqrt{g(\alpha)}d^4x f \bar{R}+\frac{\pi}{3}\frac{(1-\alpha)(1+\alpha)}{\alpha}\int_\Sigma \sqrt{\gamma} d^{2}\mathbf{x} f,\label{eq:AM_2} \\
A^{\mathcal{M}_\alpha}_2\left[f,\nabla^2\right]&=&\int_{\mathcal{M}_\alpha}\sqrt{g(\alpha)}d^4x f \left[\frac{1}{180}\bar{R}_{abcd}\bar{R}^{abcd} -\frac{1}{180}\bar{R}_{ab}\bar{R}^{ab}+\frac{1}{72}\bar{R}^2 \right]+\frac{\pi}{18}\frac{(1-\alpha)(1+\alpha)}{\alpha}\int_\Sigma \sqrt{\gamma} d^{2}\mathbf{x} f\bar{R} \nonumber  \\
&-& \frac{\pi}{180}\frac{(1-\alpha)(1+\alpha)(1+\alpha^2)}{\alpha^3}\int_\Sigma \sqrt{\gamma} d^{2}\mathbf{x} f\left[\bar{R}_{ii}-2\bar{R}_{ijij}\right], \label{eq:AM_3}
\end{eqnarray}
and
\begin{eqnarray}
   A_0\left[f\delta_\Sigma,\nabla^2\right]&=&\int_\Sigma \sqrt{\gamma} d^{2}\mathbf{x} f, \label{eq:ASigma_1}\\
   A_1\left[f\delta_\Sigma,\nabla^2\right]&=&\frac{1}{6}\int_\Sigma \sqrt{\gamma} d^{2}\mathbf{x} f\bar{R}, \label{eq:ASigma_2}\\
   A_2\left[f\delta_\Sigma,\nabla^2\right]&=&\int_{\Sigma}\sqrt{\gamma} d^2\mathbf{x} f \bigg[\frac{1}{180}\bar{R}_{abcd}\bar{R}^{abcd}-\frac{1}{180}\bar{R}_{ab}\bar{R}^{ab}+\frac{1}{72}\bar{R}^2\bigg], \label{eq:ASigma_3}
\end{eqnarray}
\end{widetext}
respectively. Here, $$\bar{R}_{ii}=\bar{R}_{ab}n^{a}_{i}n^{b}_{i},$$
and 
$$\bar{R}_{ijij}=\bar{R}_{abcd}n_{i}^{a}n_{j}^{b}n_{i}^{c}n_{j}^{d},$$ with $n_{i}^{a}$, $i=1,2$, being two orthonormal vectors to $\Sigma$. By using Eqs~(\ref{eq:AM_1})-(\ref{eq:ASigma_3})
in Eqs.~(\ref{eq:sTr_1}) and~(\ref{eq:sTr_2}) one can write the functional traces in Eq~(\ref{Eq:ERG_Flow_SGrav_expanded}) as  
\begin{eqnarray}
  D_{\alpha}\left.{\rm sTr}_{\mathcal{M}_\alpha}\left[\frac{k\partial_kR_k(\nabla^2)}{P_{k}\left(\nabla^2\right) + m_k^2}\right] \right|_{\alpha=1}\!\!\!=\frac{-A_{\Sigma}Q_1\left[W_1\left(\nabla^2\right)\right]}{24\pi}, \label{eq:sTr1_solved}\nonumber \\ 
   \\
  D_{\alpha}\left.{\rm sTr}_{\mathcal{M}_\alpha}\left[\frac{\varphi^2 k\partial_kR_k(\nabla^2)}{\left(P_{k}\left(\nabla^2\right) + m_k^2\right)^{2}}\right] \right|_{\alpha=1}\!\!\!=-\frac{Q_1\left[W_2\left(\nabla^2\right)\right]}{24\pi}\nonumber \\
   \times \int_\Sigma \sqrt{\gamma} d^{2}\mathbf{x} \varphi^2,\nonumber \\ \label{eq:sTr2_solved}
   \end{eqnarray}
\begin{eqnarray}   
   {\rm sTr}_\Sigma\left[\frac{k\partial_kR_k(\nabla^2)}{\left(P_{k}\left(\nabla^2\right) + m^2_k \right)^{2}}\right]
   &=&\frac{A_\Sigma Q_2\left[W_2\left(\nabla^2\right)\right]}{16\pi^2},\label{eq:sTr3_solved} \nonumber\\
   &&\\
  {\rm sTr}_\Sigma\left[\frac{\varphi^2 k\partial_kR_k(\nabla^2)}{\left(P_{k}\left(\nabla^2\right) + m^2_k \right)^{3}}\right]&=&\frac{Q_2\left[W_3\left(\nabla^2\right)\right]}{16\pi^2} \nonumber \\ &\times&\int_\Sigma \sqrt{\gamma} d^{2}\mathbf{x} \varphi^2.
   \label{eq:sTr4_solved}
\end{eqnarray}
To cast the functional traces in the above form, we have used that both Eq.~(\ref{eq:AM_1}) and the first term in Eqs.~(\ref{eq:AM_2}) and~(\ref{eq:AM_3}) are proportional to $\alpha$ and thus, they will vanish after we take the $D_\alpha|_{\alpha=1}$ derivative. Additionally, whenever $f=1$ or $f=\varphi^2$, the second and third terms in Eq.~(\ref{eq:AM_3}) as well as Eqs.~(\ref{eq:ASigma_2}) and~(\ref{eq:ASigma_3}) produce terms in the entropy flow which lie outside our truncation subspace. As we have pointed out, our choice of truncation submanifold for the effective action induces a truncation submanifold for the entropy.  The terms discarded in Eqs.~(\ref{eq:sTr1_solved})-(\ref{eq:sTr4_solved}) are generated by terms like 
\begin{eqnarray*}
&&c_1\int_{\mathcal{M}}\sqrt{g}d^4x R^2,\:\;\; c_2\int_{\mathcal{M}}\sqrt{g}d^4x R_{ab}R^{ab},\\
\end{eqnarray*}
\begin{eqnarray*}
&&c_3\int_{\mathcal{M}}\sqrt{g}d^4x R_{abcd}R^{abcd},\:\:\: c_4\int_{\mathcal{M}}\sqrt{g}d^4x R R_{abcd}R^{abcd},
\end{eqnarray*}
etc., in the effective action. Since such terms lie outside our truncation subspace, the contributions they generate to the entropy also lie outside the induced entropy truncation submanifold and will vanish when projected on the later~\footnote{Even if we had taken a truncation subspace that included such terms, our conclusions would be left unchanged. In this case, the quantum corrections would only renormalize the $c_j$, $j\in\mathbb{N}$, and would not mix the entanglement entropy with other entropy terms. Hence, the inclusion of such terms would only render the expressions more lengthy and would not add anything relevant to the discussion. For this reason, we neglect them without any conceptual loss.}. A completely analogous reasoning is valid for the term  $\mathcal{F}_{k}(m_{k}^{2},\lambda_{k}\varphi^{2},\xi_{k}\bar{R})$. Since all its contributions will lie outside the truncation subspace, such a term will vanish when projected on the entropy truncation subspace.  

If we now use Eqs.~(\ref{eq:sTr1_solved})-(\ref{eq:sTr4_solved}) in Eq~(\ref{Eq:ERG_Flow_SGrav_expanded}) we obtain
\begin{eqnarray}
     &&k\partial_k S^{\textrm{grav}}_k=\frac{A_{\Sigma}k^2}{8\pi}\frac{1}{\left(1+\frac{m^2_k}{k^2}\right)^2}\left[\xi_k -\frac{1}{3}\left(1+\frac{m^2_k}{k^2}\right)\right] \nonumber \\ \nonumber
     &&+ \frac{\lambda_k}{8\pi\left(1+\frac{m^2_k}{k^2}\right)^3}\left[\frac{1}{6}\left(1+\frac{m^2_k}{k^2}\right)-\xi_k\right]\int_\Sigma \sqrt{\gamma} d^{2}\mathbf{x} \; \varphi^2, \\
     \label{eq:RGSGrav3}
\end{eqnarray}  
which gives us the desired flow of the gravitational entropy~(\ref{eq:Sgrav=Sbh+Swald}) (projected onto our entropy truncation submanifold) in a spacetime with a bifurcate Killing horizon and containing a self-interacting and non-minimally coupled real scalar field.


\section{1-loop Approximation}
\label{sec:IV}

To study the consequences of ERGE~(\ref{eq:RGSGrav3}), let us make the so-called 1-loop approximation. This amounts to fix the couplings on the right-hand side of Eq.~(\ref{eq:RGSGrav3}) as $m_{k}=m_{\Lambda}$, $\xi_{k}=\xi_{\Lambda}$, and $\lambda_{k}=\lambda_{\Lambda}$, with $\Lambda$ being a UV cut-off scale where the flow begins, i.e., it begins at $k=\Lambda$ and goes to $k=0$, where all quantum fluctuations have been integrated out. As a result, we have 
\begin{eqnarray}\nonumber
      k\partial_k S^{\textrm{grav}}_k&=&-\frac{k^2}{24\pi}\frac{A_{\Sigma}}{\left(1+\frac{m^{2}_{\Lambda}}{k^{2}}\right)}+\frac{\xi_{\Lambda}k^{2}}{8\pi}\frac{A_{\Sigma}}{\left(1+\frac{m^{2}_{\Lambda}}{k^{2}}\right)^{2}}\\ \nonumber
     &&+\frac{\lambda_{\Lambda}}{48\pi\left(1+\frac{m_{\Lambda}^{2}}{k^{2}}\right)^{2}}\int_{\Sigma}\sqrt{\gamma} d^{2}\mathbf{x}\varphi^{2}\nonumber\\
     &&-\frac{\lambda_{\Lambda}\xi_{\Lambda}}{8\pi\left(1+\frac{m^{2}_{\Lambda}}{k^{2}}\right)^{3}}\int_{\Sigma}\sqrt{\gamma} d^{2}\mathbf{x}\varphi^{2}.
     \label{eq:RGSGrav4-oneloop}
\end{eqnarray}

Our first goal will be to interpret all the terms contributing to Eq.~(\ref{eq:RGSGrav4-oneloop}). To this end let us start by noting that we can write the first term on the right-hand side of Eq.~(\ref{Eq:ERG_Flow_SGrav_expanded}) in the 1-loop approximation as 
\begin{eqnarray}
     &&\frac{1}{2}D_{\alpha}\left.{\rm sTr}_{\mathcal{M}_\alpha}\left[\left(P_{k}\left(\nabla^2\right)+ m_{\Lambda}^{2}\right)^{-1}k\partial_k R_k\left(\nabla^2\right)\right]\right|_{\alpha=1}\nonumber\\ 
     &=&\left. k\partial_k \left[\frac{1}{2}D_{\alpha}\left.{\rm sTr}_{\mathcal{M}_\alpha}\log\left(\frac{P_{k}\left(\nabla^2\right)+ m_\Lambda^2}{\Lambda}\right)\right] \right|_{\alpha=1}\right.. \nonumber \\
     \label{eq:expansion_supertrace_twofirstterms_a}
\end{eqnarray}
By defining the IR effective action on $\left(\mathcal{M}_{\alpha},g_{ab}(\alpha)\right)$
\begin{align}\nonumber
    W_k(\alpha)&\equiv\frac{1}{2}{\rm sTr}_{\mathcal{M}_\alpha}\log\left(\frac{\nabla^2+ m_\Lambda^2}{\Lambda}\right)\\
    &-\frac{1}{2}{\rm sTr}_{\mathcal{M}_\alpha}\log\left(\frac{P_{k}\left(\nabla^2\right)+ m_\Lambda^2}{\Lambda}\right),
    \label{W_alpha_k}
\end{align}
which describes the field degrees of freedom outside the black hole with energy below $k^{2}$~\footnote{If $p^{2}$ are the eigenvalues of $\nabla^{2}$, as $R_k(p^2)$ vanishes for $p^2>k^2,$ it leaves the contribution of the high-energy modes to ${\rm sTr}_{\mathcal{M}_\alpha}\log\left(P_k\left(\nabla^{2}\right)+m^2_{\Lambda}\right)$ untouched while it suppresses the contribution of the low-energy modes, $p^2<k^2$. Hence, Eq.~(\ref{W_alpha_k}) gives the (low-energy) effective action at a scale $k$ (it is easy to see that $W_k$ indeed vanishes for $p^2>k^2$). A similar reasoning is valid for the other terms we analyze next.}, we have that 
\begin{equation}
    S^{\textrm{ent}}_k=\left.D_\alpha W_{k}(\alpha)\right|_{\alpha=1}
    \label{S_ent_k}
\end{equation}
is the IR entanglement entropy of the modes with energy scale below $k^2$ and thus, Eq.~(\ref{eq:expansion_supertrace_twofirstterms_a}) can be written as 
\begin{equation}
\frac{1}{2}D_{\alpha}\left.{\rm sTr}_{\mathcal{M}_\alpha}\left[\frac{k\partial_k R_k\left(\nabla^2\right)}{P_{k}\left(\nabla^2\right)+ m_{\Lambda}^{2}}\right]\right|_{\alpha=1} \!\!\!\!\!\!\!\!=-k\partial_kS_k^{\textrm{ent}}.
\label{eq:IR_entanglement_entropy_flow_oneloop}
\end{equation}
Similarly, the second term of Eq.~(\ref{Eq:ERG_Flow_SGrav_expanded}) in the 1-loop approximation can be written as 

\begin{eqnarray}
&-&\frac{\lambda_{\Lambda}}{4}D_{\alpha}\left.{\rm sTr}_{\mathcal{M}_\alpha}\left[\frac{\varphi^2k\partial_k R_k\left(\nabla^2\right)}{\left(P_{k}\left(\nabla^2\right) + m_{\Lambda}^{2}\right)^{2}}\right]\right|_{\alpha=1}\nonumber \\
&=&k\partial_k\left[ \frac{\lambda_\Lambda}{4}D_{\alpha}\left.{\rm sTr}_{\mathcal{M}_\alpha}\left(\varphi^2\left(P_{k}\left(\nabla^2\right)+ m_\Lambda^2 \right)^{-1}\right)\right] \right|_{\alpha=1}. \nonumber \\
\label{eq:expansion_supertrace_twofirstterms_b}
\end{eqnarray}
Thus, by defining the IR effective action
\begin{eqnarray}\nonumber
  W^{\lambda}_k(\alpha)&\equiv&\frac{\lambda_\Lambda }{4}{\rm sTr}_{\mathcal{M}_\alpha}\left[\varphi^2\left(\nabla^2+ m_\Lambda^2 \right)^{-1}\right]\\
  &-&\frac{\lambda_\Lambda}{4}{\rm sTr}_{\mathcal{M}_\alpha}\left[\varphi^2\left(P_{k}\left(\nabla^2\right) + m_\Lambda^2 \right)^{-1}\right],
  \label{eq:W_alpha_k_lambda}
\end{eqnarray}
  which gives the 1-loop contribution due to the self-interaction coming from low-energy modes (i.e., modes with energy scale below $k^2$) associated with the regular region in the conical spacetime, we have that 
\begin{equation}
    S_{k}^{\lambda}\equiv \left.D_{\alpha}W_{k}^{\lambda}(\alpha)\right|_{\alpha=1}.
    \label{eq:S_lambda_k}
\end{equation}
Equation~(\ref{eq:S_lambda_k}) provides the 1-loop contribution to the entropy coming from the low-energy modes due to the self-interaction. (As it will shown next, there is also an 1-loop contribution due to the self-interaction coming from the singularity at the ``tip'' of the cone.) By using Eq.~(\ref{eq:W_alpha_k_lambda}) and~(\ref{eq:S_lambda_k}) in Eq.~(\ref{eq:expansion_supertrace_twofirstterms_b}), we can write the second term on the right-hand side of Eq.~(\ref{Eq:ERG_Flow_SGrav_expanded}) as 
\begin{equation}
\frac{\lambda_{\Lambda}}{4}D_{\alpha}\left.{\rm sTr}_{\mathcal{M}_\alpha}\left[\frac{\varphi^2k\partial_k R_k\left(\nabla^2\right)}{\left(P_{k}\left(\nabla^2\right) + m_{\Lambda}^{2}\right)^{2}}\right]\right|_{\alpha=1}=k\partial_k S^{\lambda}_k. 
\label{eq:low_energy_entropy_self-interaction}
\end{equation}

The contributions coming from the functional traces over $\Sigma$ in Eq.~(\ref{Eq:ERG_Flow_SGrav_expanded}) can be treated in a similar fashion yielding
\begin{eqnarray}
&&(2\pi\xi_\Lambda){\rm sTr}_\Sigma\left[\frac{k\partial_k R_k\left(\nabla^2\right)}{\left(P_{k}\left(\nabla^2\right)+ m_\Lambda^2\right)^{2}}\right] \nonumber \\
&&=k\partial_k \left[(-2\pi\xi_\Lambda){\rm sTr}_\Sigma\left(P_{k}\left(\nabla^2\right)+m_\Lambda^2\right)^{-1}\right]
\label{eq:expansion_supertrace_twolastterms_a}
\end{eqnarray}
and
\begin{eqnarray}
&&-(2\pi\xi_\Lambda\lambda_\Lambda ){\rm sTr}_\Sigma\left[\frac{\varphi^2k\partial_k R_k\left(\nabla^2\right)}{ \left(P_{k}\left(\nabla^2\right)+m_\Lambda^2  \right)^{3}}\right]\nonumber \\
&&=k\partial_k\left[(\pi\xi_\Lambda \lambda_\Lambda){\rm sTr}_\Sigma \left(\varphi^2\left(P_{k}\left(\nabla^2\right)+m_\Lambda^2  \right)^{-2}\right)\right].
\label{eq:expansion_supertrace_twolastterms_b}
\end{eqnarray}
The right-hand side of Eq.~(\ref{eq:expansion_supertrace_twolastterms_a}) has an interesting interpretation. By noting that
\begin{equation}
{\rm sTr}_\Sigma\left(P_{k}+m_\Lambda^2  \right)^{-1}=\int_\Sigma \sqrt{\gamma} d^2{\bf x}\langle \delta \phi^2 ({\bf x})\rangle^{\textrm{free}}_{k^>},
\label{eq:trace_modified_propagator_free_theory}
\end{equation}
where the expectation value is with respect to the free theory ($\lambda_{\Lambda}=0$), gives the quantum fluctuations coming from the high-energy modes (i.e., modes with energy above $k^{2}$), we can build the following function:
\begin{eqnarray}\nonumber
&&(-2\pi\xi_\Lambda)\left[\left.{\rm sTr}_\Sigma\left(\nabla^2 + m_\Lambda^2  \right)^{-1}\right.-\left.{\rm sTr}_\Sigma\left(P_{k}+ m_\Lambda^2  \right)^{-1}\right.\right]\\
&&\equiv(-2\pi\xi_\Lambda)\int_\Sigma \sqrt{\gamma} d^2{\bf x}  \langle\delta \phi^2 ({\bf x})\rangle^{\textrm{free}}_{k^<}\nonumber \\
&&\equiv \langle S^{\textrm{wald}}_k\rangle^{\textrm{free}},
\label{eq:S_wald_free}
\end{eqnarray}
which provides the IR expectation value of Wald entropy at a scale $k$. Therefore, by using Eq.~(\ref{eq:S_wald_free}) in Eq.~(\ref{eq:expansion_supertrace_twolastterms_a}) we can write
\begin{equation}
    (2\pi\xi_\Lambda){\rm sTr}_\Sigma\left[\frac{k\partial_k R_k\left(\nabla^2\right)}{\left(P_{k}\left(\nabla^2\right)+ m_\Lambda^2\right)^{2}}\right]=-k\partial_{k}\langle S_{k}^{\textrm{wald}}\rangle^{\textrm{free}}.
    \label{eq:flow_expval_waldentropy_free}
\end{equation}
As for Eq.~(\ref{eq:expansion_supertrace_twolastterms_b}), if we define the entropy
\begin{align}\nonumber
    {}^{\Sigma}S^{\lambda}_{k}&\equiv\left(\pi\lambda_{\Lambda}\xi_{\Lambda}\right)\left.{\rm sTr}_{\Sigma}\left[\varphi^{2}\left(\nabla^{2}+m^{2}_{\Lambda}\right)^{-2}\right]\right.\\
    &-\left(\pi\lambda_{\Lambda}\xi_{\Lambda}\right)\left.{\rm sTr}_{\Sigma}\left[\varphi^{2}\left(P_{k}\left(\nabla^2\right)+m^{2}_{\Lambda}\right)^{-2}\right]\right.,
    \label{eq:S_1loop_cone_selfinteraction_singularity}
\end{align}
which will give the IR 1-loop contribution of the self-interaction to the entropy at a scale $k$ due to the conical singularity at $\Sigma$, we can cast Eq.~(\ref{eq:expansion_supertrace_twolastterms_b}) as 
\begin{equation}
-(2\pi\xi_\Lambda\lambda_\Lambda ){\rm sTr}_\Sigma\left[\frac{\varphi^2k\partial_k R_k\left(\nabla^2\right)}{ \left(P_{k}\left(\nabla^2\right)+m_\Lambda^2  \right)^{3}}\right]=-k\partial_k{}^{\Sigma}S^{\lambda}_{k}.
\label{eq:S_1loop_cone_selfinteraction_singularity_flow}
\end{equation}
This shows how one can physically interpret each term in Eq.~(\ref{Eq:ERG_Flow_SGrav_expanded}). In order to further confirm that they indeed give rise to the contributions appearing in  Eq.~(\ref{eq:RGSGrav4-oneloop}), we perform an independent calculation (whose details can be found in Appendix~\ref{sec:appa}) of the IR entropies given in Eqs.~(\ref{S_ent_k}),~(\ref{eq:S_lambda_k}),~(\ref{eq:S_wald_free}), and~(\ref{eq:S_1loop_cone_selfinteraction_singularity}). The entropy flow generated by the IR entropies are given by

\begin{eqnarray}
&& k\partial_k S^{\rm ent}_k =\frac{A_\Sigma k^2}{24\pi}\frac{1}{\left(1+ \frac{m^2_\Lambda}{k^2}\right)}, \label{1La} \\
&& k\partial_k \langle S^{\textrm{wald}}_k\rangle^{\textrm{free}} = -\frac{\xi_\Lambda  k^2}{8\pi}\frac{A_\Sigma}{\left(1+ \frac{m^2_\Lambda}{k^2}\right)^2}, \\
&& k\partial_k  S^{\lambda}_k = -\frac{\lambda_\Lambda}{48\pi}\frac{1}{\left(1+ \frac{m^2_\Lambda}{k^2}\right)^2}\int_\Sigma \sqrt{\gamma}d^2{\bf x} \varphi^2, \\
&& k\partial_k {}^\Sigma S^{\lambda}_k= \frac{\lambda_\Lambda \xi_{\Lambda}}{8\pi\left(1+ \frac{m^2_\Lambda}{k^2}\right)^3}\int_\Sigma \sqrt{\gamma}d^2{\bf x} \varphi^2. \label{1Lb}
\end{eqnarray}
As a result, by comparing Eqs.~(\ref{1La})-(\ref{1Lb}) with the right-hand side of Eq.~(\ref{eq:RGSGrav4-oneloop}), we see that we can indeed write the 1-loop RG flow of the total gravitational entropy as
\begin{eqnarray}\nonumber
k\partial_k S^{\textrm{grav}}_k&=&-k\partial_k S^{\textrm{ent}}_k-k\partial_k\langle S^{\textrm{wald}}_k\rangle^{\textrm{free}}\\
&&-k\partial_k  S^{\lambda}_k-k\partial_k {}^\Sigma S^{\lambda}_k.
  \label{eq:Gravitational_ERG_flow_interpreted}
\end{eqnarray}
The first and second term of Eq.~(\ref{eq:Gravitational_ERG_flow_interpreted}) give the flow of the entanglement entropy and Wald entropy of modes below the energy scale $k^2$, respectively. The two remaining terms come from the IR 1-loop contribution of the self-interaction to the entropy flow--the third term being the regular 1-loop contribution while the fourth term is present due to the conical singularity at $\Sigma$. 

More interestingly, we note from Eq.~(\ref{eq:Gravitational_ERG_flow_interpreted}) that the quantum contributions appearing in its right-hand side can be splitted in terms of the IR entanglement entropy flow (first term) and the IR quantum contributions to the flow of Wald entropy. As the left-hand side of Eq.~(\ref{eq:Gravitational_ERG_flow_interpreted}) can be written as 
\begin{equation}
    k\partial_{k}S^{\textrm{grav}}_{k}=k\partial_{k}S_{k}^{\textrm{BH}}+k\partial_{k}S_{k}^{\textrm{wald}},
    \label{eq:Sgrav_k}
\end{equation}
we can see that the quantum corrections preserve the split between black hole and Wald entropy throughout the ERG flow. 

If we now integrate Eq.~(\ref{eq:Gravitational_ERG_flow_interpreted}) from $k=0$ to $k$ we obtain
\begin{eqnarray}
    S^{\textrm{BH}}_0&+&S^{\textrm{wald}}_0=\left(S^{\textrm{BH}}_k+S_k^{\textrm{ent}}\right)\nonumber \\
    &+&\left(S^{\textrm{wald}}_k+\langle  S^{\textrm{wald}}_k\rangle^{\textrm{free}}+S^{\lambda}_k+{}^\Sigma S^{\lambda}_k\right), 
    \label{eq:balance}
\end{eqnarray}
where
\begin{equation}
S_k^{\textrm{ent}}=\frac{A_\Sigma }{48\pi}\left[k^2-m^2_\Lambda\log\left(\frac{k^2+ m^2_\Lambda}{m^2_\Lambda}\right)\right],\label{eq:S_ent_integrated}
\end{equation}

\begin{eqnarray} \nonumber
&&\langle S^{\textrm{wald}}_k\rangle^{\textrm{free}}=(-2\pi\xi_\Lambda)\int_\Sigma \sqrt{\gamma} d^2{\bf x}\langle\delta \phi^2 ({\bf x})\rangle^{\textrm{free}}_{k^<}\\ \nonumber
&&=\frac{\xi_\Lambda A_\Sigma}{8\pi}\left[m^2_\Lambda\log\left(\frac{k^2+ m^2_\Lambda}{m^2_\Lambda}\right)-\frac{k^2(k^2+2m^2_\Lambda)}{2(k^2+m^2_\Lambda)}\right],\\
\label{eq:S_wald_free_expval_integrated}
\end{eqnarray}

\begin{eqnarray}\nonumber
S^{\lambda}_k=\frac{\lambda_\Lambda}{96\pi}\left[\frac{k^2}{k^2+m^2_\Lambda}-\log\left(\frac{k^2+ m^2_\Lambda}{m^2_\Lambda}\right)\right]\int_\Sigma\sqrt{\gamma} d^2{\bf x} \varphi^2,\\
\label{eq:S_self-interacting_regular}
\end{eqnarray}
 
\begin{eqnarray}\nonumber
{}^\Sigma S^{\lambda}_k&=&\frac{\lambda_\Lambda \xi_{\Lambda}}{16\pi}\left[\log\left(\frac{k^2+ m^2_\Lambda}{m^2_\Lambda}\right)-\frac{k^2(3k^2+2m^2_\Lambda)}{2(k^2+m^2_\Lambda)^2}\right]\nonumber \\
&\times& \int_\Sigma \sqrt{\gamma}d^2{\bf x} \varphi^2, \label{eq:S_self-interacting_singularity}
\end{eqnarray}
and we recall that
\begin{align}
    S_{k}^{\textrm{BH}}&=\frac{A_{\Sigma}}{4G_{k}},\\
    S_{k}^{\textrm{wald}}&=-2\pi\xi_{k}\int_{\Sigma}\sqrt{\gamma} d^{2}\textbf{x}\varphi^{2}(\textbf{x}).
\end{align}
Therefore, the total (renormalized) black hole entropy, $S^{\rm BH}_{k=0}$, which can be partitioned between its effective entropy at a scale $k$, $S^{\rm BH}_k$,  and the IR entanglement entropy of scalar field modes below $k$, $S^{\rm ent}_k,$ remains constant while the balance between these two contributions changes as we slide the energy scale $k^2$. Similarly, due to the exact splitting between $S^{\rm BH}_k$ and $S^{\rm wald}_k$ for all $k$, the total (renormalized) Wald entropy  $S^{\rm wald}_{k=0}$ remains constant while the balance between its effective entropy at a scale $k$,  $S^{\rm wald}_{k},$ and the $1-$loop contributions due to modes with energy below $k
^2$ changes.

\subsection{Interacting Theories, Entropy, and Non-Minimal Coupling}

Let us now show that there is no mismatch between the renormalization of the entanglement entropy and the usual renormalization of the coupling constants coming from the effective action. In order to see this, let us use our truncation ansatz~(\ref{eq:definition_EAA_general}) into the ERGE~(\ref{Eq:ERGE}) to compute the following flow equations for $1/G_k$ and $\xi_k$ (see Appendix~\ref{appB} for the details of the calculations as well as for the flow equations of the other couplings):

\begin{equation}
k\partial_k\left(\frac{1}{G_k}\right)=\frac{k^2}{2\pi \left(1+\frac{m_k^2}{k^2}\right)^2}\left[\xi_k-\frac{1}{3}\left(1+\frac{m_k^2}{k^2}\right)\right]
\label{eq:G_flow}
\end{equation}
for Newton constant and 
\begin{equation}
k\partial_k\xi_k=\frac{\lambda_k}{16\pi^2 \left(1+\frac{m_k^2}{k^2}\right)^3}\left[\xi_k-\frac{1}{6}\left(1+\frac{m_k^2}{k^2}\right)\right] \label{eq:xi_flow}
\end{equation}
for the non-minimal coupling. Alternatively, the flow equation~(\ref{eq:RGSGrav3}) for the entropy together with the explicit expression~(\ref{Sgrav}) for the gravitational entropy will give us the following results: 
\begin{equation}
k\partial_k\left(\frac{1}{G_k}\right)=\frac{k^2}{2\pi \left(1+\frac{m_k^2}{k^2}\right)^2}\left[\xi_k-\frac{1}{3}\left(1+\frac{m_k^2}{k^2}\right)\right],\\ \label{eq:G_flow_2}
\end{equation}
and
\begin{equation}
k\partial_k\xi_k=\frac{\lambda_k}{16\pi^2 \left(1+\frac{m_k^2}{k^2}\right)^3}\left[\xi_k-\frac{1}{6}\left(1+\frac{m_k^2}{k^2}\right)\right].\\ \label{eq:xi_flow_2}
\end{equation}

Now, by comparing Eqs.~(\ref{eq:G_flow}) and~(\ref{eq:xi_flow}) with Eqs.~(\ref{eq:G_flow_2}) and~(\ref{eq:xi_flow_2}), we can see that the flow of $G_k$ and $\xi_k$ coming from the ERGE of the effective action and of the entropy match exactly. As a result, when we take the 1-loop approximation, our calculations show that the renormalization of the effective action renders the gravitational entropy automatically finite. Consequently, there is no mismatch between the renormalization of the coupling constants coming from the effective action or the gravitational entropy. 

The origin of the ``puzzle'' is the absence of the last term in~(\ref{eq:RGSGrav4-oneloop}) [which is highlighted in Eq.~(\ref{1Lb})] in the usual 1-loop calculations for the entanglement entropy of a self-interacting scalar field~\cite{Solodukhin_nonrelativistic}. However, we see that it appears quite naturally when one uses the ERGE~(\ref{Eq:ERGE}) to derive the flow of the total gravitational entropy in a consistent way. Thus, the concern that the renormalization of $\xi$ coming from the effective action would not guarantee that the total entropy $S^{\textrm{grav}}_{\Lambda}+ S^{\textrm{1-loop}}_{k=\Lambda}$ would be finite (and therefore would pose an additional problem in interpreting the entanglement entropy as the origin of black hole entropy) is unjustified.


\section{Conclusions}\label{sec:V}

We have considered a non-minimally coupled and self-interacting quantum scalar field $\phi$ in a spacetime with a bifurcate Killing horizon $\mathfrak{h}$ and used the ERGE to find the RG flow of the total gravitational entropy $$S^{\rm grav}_k=S^{\rm BH}_k + S^{\rm wald}_k$$ when the IR scale $k$ is varied. We have shown that, in the 1-loop approximation, the contribution of $\phi$ to the total entropy can be splitted in terms of the entanglement entropy of modes below the scale $k$ and the quantum contributions of modes below $k$ to Wald entropy. In particular, the integrated flow makes it clear that, as $k$ is varied, the total (renormalized) black hole entropy remains constant while the balance between the contribution coming from the effective black hole entropy at a scale $k$ and the IR entanglement entropy changes. A similar conclusion is valid for the total (renormalized) Wald entropy. As $k$ is varied, the balance between the effective entropy at a scale $k$ and its IR quantum corrections changes while keeping the total renormalized entropy constant.

In addition, we have shown that the RG flow of the coupling constants coming from the effective action and from the entropy match exactly. In particular, when we take the 1-loop approximation, the renormalization of the effective action renders the total entropy   $S^{\textrm{grav}}_{\Lambda}+ S^{\textrm{1-loop}}_{k=\Lambda}$ finite. This solves an apparent ``puzzle" that appeared to exist for interacting fields.

In order to answer whether or not all black hole entropy can be  interpreted as entanglement entropy depends on knowing if gravity at low energy is completely ``induced" by integrating out all quantum field fluctuations~\cite{S68, J94a,J95} or if there is a UV complete quantum gravity theory. Notwithstanding this, our results show that even for a non-minimally coupled and self-interacting scalar field, the entanglement entropy of modes with energy scale below $k^2$ outside the horizon can be seen as responsible for generating at least part of the black hole entropy.


\begin{appendix}
\section{IR Contributions to the Effective Action and their Entropy Flow}
\label{sec:appa}
In this appendix, we perform an independent calculation of the RG flow of the IR entropies $S^{\textrm{ent}}_{k}$, $S^{\lambda}_{k}$, $\langle  S^{\textrm{wald}}_k\rangle^{\textrm{free}},$ and ${}^\Sigma S^{\lambda}_k$. Such calculation will not only show the expected agreement with each term of Eq.~(\ref{eq:RGSGrav4-oneloop}) but also will allow us to interpret them physically as we have done in the main text.  

In order to compute the functional traces appearing in Eqs.~(\ref{S_ent_k}),~(\ref{eq:S_lambda_k}),~(\ref{eq:S_wald_free}), and~(\ref{eq:S_1loop_cone_selfinteraction_singularity}), we first use the  Heat Kernel method to write
\begin{widetext}

\begin{eqnarray}
   \frac{1}{2}\textrm{sTr}_{\mathcal{M}_{\alpha}}\log\left(\frac{P_{k}(\nabla^{2})+m^{2}_{\Lambda}}{\Lambda}\right)&=&\frac{1}{(4\pi)^{2}}\sum_{k=0}^{\infty}Q_{2-k}\left[\hat{W}(\nabla^{2})\right]A_{k}^{\mathcal{M}_{\alpha}}\left[\nabla^{2}\right]\label{eq:sTr_1_oneloopsection},\\ 
   \textrm{sTr}_{\mathcal{M}_{\alpha}}\left[f\left(P_{k}(\nabla^{2})+m^{2}_{\Lambda}\right)^{-l}\right]&=&\frac{1}{(4\pi)^{2}}\sum_{k=0}^{\infty}Q_{2-k}\left[\tilde{F}_{l}(\nabla^{2})\right]A_{k}^{\mathcal{M}_{\alpha}}\left[f,\nabla^{2}\right]\label{eq:sTr_2_oneloopsection},\\
   \textrm{sTr}_{\Sigma}\left[f\left(P_{k}(\nabla^{2})+m^{2}_{\Lambda}\right)^{-l}\right]&=&\frac{1}{(4\pi)^{2}}\sum_{k=0}^{\infty}Q_{2-k}\left[\tilde{F}_{l}(\nabla^{2})\right]A_{k}\left[f\delta_{\Sigma},\nabla^{2}\right]\label{eq:sTr_3_oneloopsection},
\end{eqnarray}
\end{widetext}
where $l=1,2$, $f$ is a scalar function, $\hat{W}$ has the form
\begin{equation}
    \hat{W}(z)\equiv\frac{1}{2}\log\left[\frac{k^{2}}{\Lambda}\left(z+\frac{m^{2}_{\Lambda}}{\Lambda}+R^{(0)}(z)\right)\right]
    \label{eq:hat_W_function},
\end{equation}
and we have defined $\tilde{F}_{l}(p^{2})\equiv k^{-2l}F_{l}(z)$, with
\begin{equation}
    F_{l}(z)=\left(z+\frac{m^{2}_{\Lambda}}{k^{2}}+R^{(0)}(z)\right)^{-l}
    \label{eq:F_function}
\end{equation}
and $R^{(0)}(z)=(1-z)\theta(1-z)$. By using  Eqs.~(\ref{eq:AM_1})-(\ref{eq:ASigma_3}) to evaluate the supertraces in Eqs.~(\ref{eq:sTr_1_oneloopsection})-(\ref{eq:sTr_3_oneloopsection}) and projecting onto our entropy truncation subspace we obtain 
\begin{eqnarray}\nonumber
&&\frac{1}{2}D_{\alpha}\left. {\rm sTr}_{\mathcal{M}_\alpha}\log\left(\frac{P_{k}\left(\nabla^2\right) + m_\Lambda^2}{\Lambda}\right)\right|_{\alpha=1}\nonumber \\
&=&-\frac{A_\Sigma Q_{1}\left[\hat{W}(\nabla^2)\right]}{24\pi},
\label{eq:sTr_F1_a}\\ \nonumber
&&\frac{1}{2}D_{\alpha}\left.{\rm sTr}_{\mathcal{M}_\alpha}\left[f\left(P_{k}\left(\nabla^2\right) + m^2_\Lambda\right)^{-l}\right]\right|_{\alpha=1}\\
&=&-\frac{Q_{1}\left[\tilde{F}_l(\nabla^2)\right]}{48\pi} \int_\Sigma \sqrt{\gamma} d^2{\bf x}f,
\label{eq:sTr_F2_b}\\ \nonumber
&&{\rm sTr}_\Sigma\left[f\left(P_{k}\left(\nabla^2\right) + m^2_\Lambda\right)^{-l}\right]\\
&=&\frac{Q_{2}\left[\tilde{F}_l(\nabla^2)\right]}{16\pi^2} \int_\Sigma \sqrt{\gamma} d^2{\bf x}f,
\label{eq:sTr_F3_b}
\end{eqnarray}
\begin{widetext}
where the Q-functionals present in these three expressions are given by
\begin{align}\nonumber
&Q_{1}\left[\hat{W}(\nabla^2)\right]=k^2\int_{0}^{\infty} dz \hat{W}(z)=\frac{k^2}{2}\log\left(\frac{k^2+m^2_\Lambda}{\Lambda}\right) + \frac{1}{2}\left(\Lambda +m^2_\Lambda\right) \log\left(\frac{\Lambda +m^2_\Lambda}{\Lambda}\right)\\
&\:\:\:\:\:\:\:\:\:\:\:\:\:\:\:\:\:\:\:\:\:\:\:\:\:\:\:\:\:\:\:\:\:\:\:\:\:\:\:\:\:\:\:\:\:\:\:\:\:\:\:\:\:\:\:\:\:\:\:\:\:-\frac{\Lambda}{2}+\frac{k^2}{2}-\frac{1}{2}\left(k^2+ m^2_\Lambda\right)\log\left(\frac{k^2+m^2_\Lambda}{\Lambda}\right)\label{eq:Q1(W)},
\\ \nonumber
\\
&Q_1\left[\tilde{F}_l\left(\nabla^2\right)\right]=k^{2(1-l)}\int_0^\infty dz F_l(z) =\left\{
\begin{array}{cc} 
\left(1+\frac{m_\Lambda^2}{k^2}\right)^{-1}+ \log\left(\frac{\Lambda + m^2_\Lambda}{k^2+ m^2_\Lambda}\right), & l=1  \\
\\
\frac{1}{k^{2}}\left[\frac{2+ m^2_\Lambda/k^2}{\left(1+ m^2_\Lambda/k^2\right)^2}\right],  & l=2
\end{array}
\right. ,
\label{eq:Q1(F)}\\ \nonumber
\\
&Q_2\left[\tilde{F}_l\left(\nabla^2\right)\right]=k^{2(2-l)}\int_0^\infty dz F_l(z)=\left\{
\begin{array}{cc} 
-\frac{k^4+2m^2_\Lambda k^2}{2(k^2 + m^2_\Lambda)}- m^2_\Lambda\log\left(\frac{\Lambda + m^2_\Lambda}{k^2+ m^2_\Lambda}\right) + \Lambda, & l=1  \\
\\
\frac{k^4}{2(k^2+ m^2_\Lambda)^2}+\frac{m^2_\Lambda\left(k^2-\Lambda\right)}{(k^2+ m^2_\Lambda)(\Lambda+ m^2_\Lambda)} + \log\left(\frac{\Lambda + m^2_\Lambda}{k^2+ m^2_\Lambda}\right),  & l=2
\end{array}
\right., \label{eq:Q2(F)}
\end{align}
and we note that whenever one of the integrals above diverge, we have imposed a cut-off, $\Lambda/k^{2},$ in its upper limit. 
\end{widetext}

Now, by applying $k\partial_{k}$ to Eq.~(\ref{eq:sTr_F1_a}) and using Eq.~(\ref{S_ent_k}) we obtain
\begin{equation}
    k\partial_k S^{\textrm{ent}}_k=\frac{A_\Sigma k^2}{24\pi}\frac{1}{\left(1+ \frac{m^2_\Lambda}{k^2}\right)}, \label{eq:1-loop_entanglemententropy_flow}
\end{equation}
which corresponds to the first term on the right-hand side Eq.~(\ref{eq:RGSGrav4-oneloop}). Similarly, by taking $f=\varphi^2$ and applying $k\partial_{k}$ on  Eq.~(\ref{eq:sTr_F3_b}) with $l=1$ and using Eq.~(\ref{eq:S_wald_free}) we obtain 
\begin{equation}
    k\partial_k \langle S^{\textrm{wald}}_k\rangle ^{\textrm{free}}= -\frac{\xi_\Lambda  k^2}{8\pi}\frac{A_\Sigma}{\left(1+ \frac{m^2_\Lambda}{k^2}\right)^2},
    \label{eq:1-loop_IRexpvalwaldentropy_flow}
\end{equation}
giving us the second term in Eq~(\ref{eq:RGSGrav4-oneloop}). Finally, by setting $f=\varphi^2$, applying  $k\partial_{k}$ to Eq.~(\ref{eq:sTr_F2_b}) with  $l=1$ and $l=2$, and making use of Eqs.~(\ref{eq:S_lambda_k}) and~(\ref{eq:S_1loop_cone_selfinteraction_singularity}), we obtain

\begin{equation}
    k\partial_k  S^{\lambda}_k=-\frac{\lambda_\Lambda}{48\pi}\frac{1}{\left(1+ \frac{m^2_\Lambda}{k^2}\right)^2}\int_\Sigma \sqrt{\gamma} d^2{\bf x}\varphi^2,
    \label{eq:1-loop_selfinteraction_regular}
\end{equation}
and
\begin{equation}
    k\partial_k {}^\Sigma S^{\lambda}_k= \frac{\lambda_\Lambda \xi_{\Lambda}}{8\pi\left(1+ \frac{m^2_\Lambda}{k^2}\right)^3}\int_\Sigma \sqrt{\gamma} d^2{\bf x} \varphi^2.
    \label{eq:1-loop_selfinteraction_singularity}
\end{equation}
\vspace{0.09cm}\noindent This gives us the remaining terms present in the right-hand side of Eq.~(\ref{eq:RGSGrav4-oneloop}).

\section{ERG Flow of the Coupling Constants}
\label{appB}
In order to compute the ERG flow of the coupling constants $G_k$, $\Lambda_k$, $m_k$, $\xi_k,$ and $\lambda_k$, we will use the ERGE
\begin{equation}
    k\partial_k\Gamma_k=\frac{1}{2}{\rm sTr}\left[\left(\Gamma^{(2)}_k + R_k(\nabla^2)\right)^{-1}k\partial_k R_k(\nabla^2)\right],
    \label{eq:ERGE_appendix}
\end{equation}
together with our EAA truncation
\begin{equation}
\Gamma_k[g_{ab},\varphi]=\Gamma_{k}^{\textrm{EH}}[g_{ab}] + \Gamma_k^\phi[g_{ab},\varphi] + \Gamma^{\textrm{int}}_k[g_{ab},\varphi],
\label{eq:EAA_appendix}
\end{equation}
where $\Gamma_{k}^{\textrm{EH}}[g_{ab}]$, $\Gamma_k^\phi[g_{ab},\varphi] $, and $\Gamma^{\textrm{int}}_k[g_{ab},\varphi]$ are given by Eqs.~(\ref{eq:Einstein-Hilbert action}),~(\ref{eq:free_scalar_field_action}), and~(\ref{eq:interaction_part_scalar_field}), respectively, and 
\begin{equation}
    \Gamma^{(2)}_k[g_{ab}, \varphi]+ R_k(\nabla^2)\equiv P_{k}\left(\nabla^2\right)+m^2_k+r_k ,
\end{equation}
whith $P_{k}\left(\nabla^2\right)\equiv\nabla^2 + R_k(\nabla^2)$ and $r_k\equiv\xi_k R+ \lambda_k\varphi^2/2.$
To be consistent with our approximations, we will need to project the right-hand side of Eq.~(\ref{eq:ERGE_appendix}) onto our truncation subspace. In order to do this, let us expand it in powers of $r_k$ discarding terms of order $\mathcal{O}\left(\lambda_k^3\varphi^6, R_{abcd}^2\right)$ or higher as

\begin{widetext}  
\begin{eqnarray}
k\partial_k \Gamma_k&=&\frac{1}{2}{\rm sTr}\left[\left( P_{k}\left(\nabla^2\right)+m^2_k \right)^{-1}k\partial_k R_k\left(\nabla^2\right)\right]-\frac{1}{2}{\rm sTr}\left[\left( P_{k}\left(\nabla^2\right)+m^2_k \right)^{-2}k\partial_k R_k\left(\nabla^2\right)\left(\xi_k R+ \frac{1}{2}\lambda_k\varphi^2\right)\right]\nonumber\\ 
&+&\frac{1}{2}{\rm sTr}\left[\left( P_{k}\left(\nabla^2\right)+m^2_k \right)^{-3}k\partial_k R_k\left(\nabla^2\right)\left(\lambda_k\xi_k R\varphi^2+ \frac{1}{4}\lambda_k^2\varphi^4\right)\right]+\mathcal{O}\left(\lambda_k^3\varphi^6, R_{abcd}^2\right).
\label{eq:ERGE_developed_appendix}
\end{eqnarray}

In order to compute the functional traces in the right-hand side of the above equation, we will make use Heat Kernel expansion of $\nabla^2$ to write

\begin{eqnarray}
&&{\rm sTr}\left[\frac{f k\partial_k R_k(\nabla^2)}{\left(P_{k}\left(\nabla^2\right) + m_k\right)^l}\right] =\frac{1}{(4\pi)^2}\sum_{k=0}^\infty Q_{2-k}\left[W_l(\nabla^2)\right]A^{\mathcal{M}}_k\left[f,\nabla^2\right],\label{eq:str1_appendix}
\end{eqnarray}
where $f$ is a scalar function on $\mathcal{M}$, 
\begin{eqnarray}
   A_0\left[f,\nabla^2\right]&=&\int_\mathcal{M} \sqrt{g}d^4{ x} f, \label{eq:A0_appendix}\\
    A_1\left[f,\nabla^2\right]&=&\frac{1}{6}\int_\mathcal{M} \sqrt{g}d^4{ x} f R, \label{eq:A1_appendix}\\
    A_2\left[f,\nabla^2\right]&=&\int_\mathcal{M} \sqrt{g}d^4{ x}f  \left[\frac{1}{180}R_{abcd}R^{abcd}-\frac{1}{180}R_{ab}R^{ab}+\frac{1}{72}R^2 \right], \label{eq:A3_appendix}
\end{eqnarray}
and
\begin{equation}
Q_n\left[W_l(\nabla^2)\right]=\left\{
\begin{array}{cc} 
2k^{2(n-l+1)}/n!(1+\frac{q_k}{k^2})^l,  & n\geq 0  \\
0,  & n<0   
\end{array}
\right. ,
\label{eq:Q-functional_appendix}
\end{equation}
\end{widetext}
with $q_{k}\equiv m^{2}_{k}+\lambda_{k}\varphi^{2}/2$. 

As a result, by using Eqs.~(\ref{eq:str1_appendix})-(\ref{eq:A3_appendix}) in Eq.~(\ref{eq:ERGE_developed_appendix}) and projecting onto the truncation submanifold we obtain 
\begin{widetext}
\begin{eqnarray}
k\partial_k\Gamma_k&=&\frac{1}{32\pi^2}Q_2\left[W_1(\nabla^2)\right]\int_\mathcal{M}\sqrt{g}d^4x-\frac{\lambda_k }{64\pi^2}Q_2\left[W_2(\nabla^2)\right]\int_\mathcal{M}\sqrt{g} d^4x\varphi^2 \nonumber \\
&+&\frac{1}{32\pi^2}\left(\frac{1}{6}Q_1\left[W_1(\nabla^2)\right]-Q_2\left[W_2(\nabla^2)\right]\right)\int_\mathcal{M}\sqrt{g}d^4x R\nonumber \\
&+&\frac{\lambda_k}{32\pi^2}\left(\xi_k Q_2\left[W_3(\nabla^2)\right]-\frac{1}{12}Q_1\left[W_1(\nabla^2)\right]\right)\int_\mathcal{M}\sqrt{g}d^4x R \varphi^2\nonumber \\
&+&\frac{\lambda_k^2 }{128\pi^2}Q_2\left[W_3(\nabla^2)\right]\int_\mathcal{M}\sqrt{g}d^4x\varphi^4 +  \mathcal{O}\left(\lambda_k^3\varphi^6, R_{abcd}^2\right).
\end{eqnarray}
If we now use Eq.~(\ref{eq:EAA_appendix}) on  left-hand side of the above equation and Eq.~(\ref{eq:Q-functional_appendix}) on the right-hand side we obtain 
\end{widetext}
\begin{eqnarray}
&&k\partial_k\left(\frac{1}{G_k}\right)=\frac{k^2}{2\pi \left(1+\frac{m_k^2}{k^2}\right)^2}\left[\xi_k-\frac{1}{3}\left(1+\frac{m_k^2}{k^2}\right)\right], \nonumber \\
\\
&&k\partial_k\left(\frac{\Lambda_k}{G_k}\right)=\frac{k^4}{4\pi \left(1+\frac{m_k^2}{k^2}\right)},\\
&&k\partial_k m^2_k =-\frac{\lambda_k k^2 }{32\pi^2 \left(1+\frac{m_k^2}{k^2}\right)^2},\\
&&k\partial_k\lambda_k=\frac{3\lambda_k^2}{16\pi^2 \left(1+\frac{m_k^2}{k^2}\right)^3},\\ \nonumber
&&k\partial_k\xi_k=\frac{\lambda_k}{16\pi^2 \left(1+\frac{m_k^2}{k^2}\right)^3}\left[\xi_k-\frac{1}{6}\left(1+\frac{m_k^2}{k^2}\right)\right],\\
\end{eqnarray}
for the RG flow of Newton constant, cosmological constant, field mass, interaction coupling, and non-minimal coupling, respectively.

\end{appendix}

\acknowledgments
A. L. and J.M. were partially and fully supported by S\~ao Paulo Research Foundation (FAPESP) under grant 2017/15084-6   and Coordena\c{c}\~ao de Aperfei\c{c}oamento de Pessoal de N\'ivel Superior (Capes) under grant 88882.451665/2019-01, respectively.

\bibliography{bibliography}

\begin{thebibliography}{37}%
\makeatletter
\providecommand \@ifxundefined [1]{%
 \@ifx{#1\undefined}
}%
\providecommand \@ifnum [1]{%
 \ifnum #1\expandafter \@firstoftwo
 \else \expandafter \@secondoftwo
 \fi
}%
\providecommand \@ifx [1]{%
 \ifx #1\expandafter \@firstoftwo
 \else \expandafter \@secondoftwo
 \fi
}%
\providecommand \natexlab [1]{#1}%
\providecommand \enquote  [1]{``#1''}%
\providecommand \bibnamefont  [1]{#1}%
\providecommand \bibfnamefont [1]{#1}%
\providecommand \citenamefont [1]{#1}%
\providecommand \href@noop [0]{\@secondoftwo}%
\providecommand \href [0]{\begingroup \@sanitize@url \@href}%
\providecommand \@href[1]{\@@startlink{#1}\@@href}%
\providecommand \@@href[1]{\endgroup#1\@@endlink}%
\providecommand \@sanitize@url [0]{\catcode `\\12\catcode `\$12\catcode
  `\&12\catcode `\#12\catcode `\^12\catcode `\_12\catcode `\%12\relax}%
\providecommand \@@startlink[1]{}%
\providecommand \@@endlink[0]{}%
\providecommand \url  [0]{\begingroup\@sanitize@url \@url }%
\providecommand \@url [1]{\endgroup\@href {#1}{\urlprefix }}%
\providecommand \urlprefix  [0]{URL }%
\providecommand \Eprint [0]{\href }%
\providecommand \doibase [0]{https://doi.org/}%
\providecommand \selectlanguage [0]{\@gobble}%
\providecommand \bibinfo  [0]{\@secondoftwo}%
\providecommand \bibfield  [0]{\@secondoftwo}%
\providecommand \translation [1]{[#1]}%
\providecommand \BibitemOpen [0]{}%
\providecommand \bibitemStop [0]{}%
\providecommand \bibitemNoStop [0]{.\EOS\space}%
\providecommand \EOS [0]{\spacefactor3000\relax}%
\providecommand \BibitemShut  [1]{\csname bibitem#1\endcsname}%
\let\auto@bib@innerbib\@empty
\bibitem [{\citenamefont {Hawking}(1975)}]{hawking1975}%
  \BibitemOpen
  \bibfield  {author} {\bibinfo {author} {\bibfnamefont {S.~W.}\ \bibnamefont
  {Hawking}},\ }\bibfield  {title} {\bibinfo {title} {Particle creation by
  black holes},\ }\href {https://projecteuclid.org:443/euclid.cmp/1103899181}
  {\bibfield  {journal} {\bibinfo  {journal} {Comm. Math. Phys.}\ }\textbf
  {\bibinfo {volume} {43}},\ \bibinfo {pages} {199} (\bibinfo {year}
  {1975})}\BibitemShut {NoStop}%
\bibitem [{\citenamefont {Wald}(1984)}]{wald84}%
  \BibitemOpen
  \bibfield  {author} {\bibinfo {author} {\bibfnamefont {R.~M.~W.}\
  \bibnamefont {Wald}},\ }\href {https://doi.org/10.1017/9781316227596} {\emph
  {\bibinfo {title} {General Relativity}}}\ (\bibinfo  {publisher} {The
  University of Chicago Press},\ \bibinfo {year} {1984})\BibitemShut {NoStop}%
\bibitem [{\citenamefont
  {Carlip}(2014)}]{Carlip_blackholethermodynamics_review}%
  \BibitemOpen
  \bibfield  {author} {\bibinfo {author} {\bibfnamefont {S.}~\bibnamefont
  {Carlip}},\ }\bibfield  {title} {\bibinfo {title} {{Black Hole
  Thermodynamics}},\ }\href {https://doi.org/10.1142/S0218271814300237}
  {\bibfield  {journal} {\bibinfo  {journal} {Int. J. Mod. Phys. D}\ }\textbf
  {\bibinfo {volume} {23}},\ \bibinfo {pages} {1430023} (\bibinfo {year}
  {2014})},\ \Eprint {https://arxiv.org/abs/gr-qc/1410.1486}
  {arXiv:gr-qc/1410.1486} \BibitemShut {NoStop}%
\bibitem [{\citenamefont
  {Wall}(2018)}]{wall_blackholethermodynamics_reviewsurvey}%
  \BibitemOpen
  \bibfield  {author} {\bibinfo {author} {\bibfnamefont {A.~C.}\ \bibnamefont
  {Wall}},\ }\href@noop {} {\bibinfo {title} {A survey of black hole
  thermodynamics}} (\bibinfo {year} {2018}),\ \Eprint
  {https://arxiv.org/abs/gr-qc/1804.10610} {arXiv:gr-qc/1804.10610}
  \BibitemShut {NoStop}%
\bibitem [{\citenamefont {Solodukhin}(2011)}]{Solodukhin_review}%
  \BibitemOpen
  \bibfield  {author} {\bibinfo {author} {\bibfnamefont {S.~N.}\ \bibnamefont
  {Solodukhin}},\ }\bibfield  {title} {\bibinfo {title} {{Entanglement entropy
  of black holes}},\ }\href {https://doi.org/10.12942/lrr-2011-8} {\bibfield
  {journal} {\bibinfo  {journal} {Living Rev. Rel.}\ }\textbf {\bibinfo
  {volume} {14}},\ \bibinfo {pages} {8} (\bibinfo {year} {2011})},\ \Eprint
  {https://arxiv.org/abs/hep-th/1104.3712} {arXiv:hep-th/1104.3712}
  \BibitemShut {NoStop}%
\bibitem [{\citenamefont
  {Solodukhin}(1995{\natexlab{a}})}]{Solodukhin-conicalsingularity}%
  \BibitemOpen
  \bibfield  {author} {\bibinfo {author} {\bibfnamefont {S.~N.}\ \bibnamefont
  {Solodukhin}},\ }\bibfield  {title} {\bibinfo {title} {Conical singularity
  and quantum corrections to the entropy of a black hole},\ }\href
  {https://doi.org/10.1103/PhysRevD.51.609} {\bibfield  {journal} {\bibinfo
  {journal} {Phys. Rev. D}\ }\textbf {\bibinfo {volume} {51}},\ \bibinfo
  {pages} {609} (\bibinfo {year} {1995}{\natexlab{a}})},\ \Eprint
  {https://arxiv.org/abs/hep-th/9407001} {arXiv:hep-th/9407001} \BibitemShut
  {NoStop}%
\bibitem [{\citenamefont {Gibbons}\ and\ \citenamefont
  {Hawking}(1977)}]{gibbons_hawking_euclidean}%
  \BibitemOpen
  \bibfield  {author} {\bibinfo {author} {\bibfnamefont {G.~W.}\ \bibnamefont
  {Gibbons}}\ and\ \bibinfo {author} {\bibfnamefont {S.~W.}\ \bibnamefont
  {Hawking}},\ }\bibfield  {title} {\bibinfo {title} {Action integrals and
  partition functions in quantum gravity},\ }\href
  {https://doi.org/10.1103/PhysRevD.15.2752} {\bibfield  {journal} {\bibinfo
  {journal} {Phys. Rev. D}\ }\textbf {\bibinfo {volume} {15}},\ \bibinfo
  {pages} {2752} (\bibinfo {year} {1977})}\BibitemShut {NoStop}%
\bibitem [{\citenamefont {Callan}\ and\ \citenamefont
  {Wilczek}(1994)}]{Callan_1994}%
  \BibitemOpen
  \bibfield  {author} {\bibinfo {author} {\bibfnamefont {J.}~\bibnamefont
  {Callan}, \bibfnamefont {Curtis~G.}}\ and\ \bibinfo {author} {\bibfnamefont
  {F.}~\bibnamefont {Wilczek}},\ }\bibfield  {title} {\bibinfo {title} {{On
  geometric entropy}},\ }\href {https://doi.org/10.1016/0370-2693(94)91007-3}
  {\bibfield  {journal} {\bibinfo  {journal} {Phys. Lett. B}\ }\textbf
  {\bibinfo {volume} {333}},\ \bibinfo {pages} {55} (\bibinfo {year} {1994})},\
  \Eprint {https://arxiv.org/abs/hep-th/9401072} {arXiv:hep-th/9401072}
  \BibitemShut {NoStop}%
\bibitem [{\citenamefont {Kabat}(1996)}]{kabat}%
  \BibitemOpen
  \bibfield  {author} {\bibinfo {author} {\bibfnamefont {D.~N.}\ \bibnamefont
  {Kabat}},\ }\bibfield  {title} {\bibinfo {title} {Black hole entropy and
  entropy of entanglement},\ }\href
  {https://doi.org/10.1016/0550-3213(95)00443-V} {\bibfield  {journal}
  {\bibinfo  {journal} {Nucl. Phys. B}\ }\textbf {\bibinfo {volume} {453}},\
  \bibinfo {pages} {281} (\bibinfo {year} {1996})},\ \Eprint
  {https://arxiv.org/abs/hep-th/9503016} {arXiv:hep-th/9503016} \BibitemShut
  {NoStop}%
\bibitem [{\citenamefont {Solodukhin}(2012)}]{solodukhin12}%
  \BibitemOpen
  \bibfield  {author} {\bibinfo {author} {\bibfnamefont {S.~N.}\ \bibnamefont
  {Solodukhin}},\ }\bibfield  {title} {\bibinfo {title} {{Remarks on effective
  action and entanglement entropy of Maxwell field in generic gauge}},\ }\href
  {https://doi.org/10.1007/JHEP12(2012)036} {\bibfield  {journal} {\bibinfo
  {journal} {J. High Energ. Phys.}\ }\textbf {\bibinfo {volume} {12}},\
  \bibinfo {pages} {36}},\ \Eprint {https://arxiv.org/abs/hep-th/1209.2677}
  {arXiv:hep-th/1209.2677} \BibitemShut {NoStop}%
\bibitem [{\citenamefont {Donnelly}\ and\ \citenamefont
  {Wall}(2012)}]{donnelly_wall}%
  \BibitemOpen
  \bibfield  {author} {\bibinfo {author} {\bibfnamefont {W.}~\bibnamefont
  {Donnelly}}\ and\ \bibinfo {author} {\bibfnamefont {A.~C.}\ \bibnamefont
  {Wall}},\ }\bibfield  {title} {\bibinfo {title} {Do gauge fields really
  contribute negatively to black hole entropy?},\ }\href
  {https://doi.org/10.1103/PhysRevD.86.064042} {\bibfield  {journal} {\bibinfo
  {journal} {Phys. Rev. D}\ }\textbf {\bibinfo {volume} {86}},\ \bibinfo
  {pages} {064042} (\bibinfo {year} {2012})},\ \Eprint
  {https://arxiv.org/abs/hep-th/1206.5831} {arXiv:hep-th/1206.5831}
  \BibitemShut {NoStop}%
\bibitem [{\citenamefont
  {Solodukhin}(1995{\natexlab{b}})}]{Solodukhin_oneloop}%
  \BibitemOpen
  \bibfield  {author} {\bibinfo {author} {\bibfnamefont {S.~N.}\ \bibnamefont
  {Solodukhin}},\ }\bibfield  {title} {\bibinfo {title} {One-loop
  renormalization of black hole entropy due to nonminimally coupled matter},\
  }\href {https://doi.org/10.1103/PhysRevD.52.7046} {\bibfield  {journal}
  {\bibinfo  {journal} {Phys. Rev. D}\ }\textbf {\bibinfo {volume} {52}},\
  \bibinfo {pages} {7046} (\bibinfo {year} {1995}{\natexlab{b}})},\ \Eprint
  {https://arxiv.org/abs/hep-th/9504022} {arXiv:hep-th/9504022} \BibitemShut
  {NoStop}%
\bibitem [{\citenamefont {Wald}(1993)}]{waldentropy}%
  \BibitemOpen
  \bibfield  {author} {\bibinfo {author} {\bibfnamefont {R.~M.}\ \bibnamefont
  {Wald}},\ }\bibfield  {title} {\bibinfo {title} {Black hole entropy is the
  {Noether} charge},\ }\href {https://doi.org/10.1103/PhysRevD.48.R3427}
  {\bibfield  {journal} {\bibinfo  {journal} {Phys. Rev. D}\ }\textbf {\bibinfo
  {volume} {48}},\ \bibinfo {pages} {R3427} (\bibinfo {year} {1993})},\ \Eprint
  {https://arxiv.org/abs/gr-qc/9307038} {arXiv:gr-qc/9307038} \BibitemShut
  {NoStop}%
\bibitem [{\citenamefont {Solodukhin}(2010)}]{Solodukhin_nonrelativistic}%
  \BibitemOpen
  \bibfield  {author} {\bibinfo {author} {\bibfnamefont {S.~N.}\ \bibnamefont
  {Solodukhin}},\ }\bibfield  {title} {\bibinfo {title} {{Entanglement Entropy
  in Non-Relativistic Field Theories}},\ }\href
  {https://doi.org/10.1007/JHEP04(2010)101} {\bibfield  {journal} {\bibinfo
  {journal} {JHEP}\ }\textbf {\bibinfo {volume} {04}},\ \bibinfo {pages}
  {101}},\ \Eprint {https://arxiv.org/abs/hep-th/0909.0277}
  {arXiv:hep-th/0909.0277} \BibitemShut {NoStop}%
\bibitem [{\citenamefont {Nelson}\ and\ \citenamefont
  {Panangaden}(1982)}]{NP82}%
  \BibitemOpen
  \bibfield  {author} {\bibinfo {author} {\bibfnamefont {B.~L.}\ \bibnamefont
  {Nelson}}\ and\ \bibinfo {author} {\bibfnamefont {P.}~\bibnamefont
  {Panangaden}},\ }\bibfield  {title} {\bibinfo {title} {Scaling behavior of
  interacting quantum fields in curved space-time},\ }\href
  {https://doi.org/10.1103/PhysRevD.25.1019} {\bibfield  {journal} {\bibinfo
  {journal} {Phys. Rev. D}\ }\textbf {\bibinfo {volume} {25}},\ \bibinfo
  {pages} {1019} (\bibinfo {year} {1982})}\BibitemShut {NoStop}%
\bibitem [{\citenamefont {Wetterich}(1993)}]{Wetterich_1993}%
  \BibitemOpen
  \bibfield  {author} {\bibinfo {author} {\bibfnamefont {C.}~\bibnamefont
  {Wetterich}},\ }\bibfield  {title} {\bibinfo {title} {{Exact evolution
  equation for the effective potential}},\ }\href
  {https://doi.org/10.1016/0370-2693(93)90726-X} {\bibfield  {journal}
  {\bibinfo  {journal} {Phys. Lett. B}\ }\textbf {\bibinfo {volume} {301}},\
  \bibinfo {pages} {90} (\bibinfo {year} {1993})},\ \Eprint
  {https://arxiv.org/abs/hep-th/1710.05815} {arXiv:hep-th/1710.05815}
  \BibitemShut {NoStop}%
\bibitem [{\citenamefont {Reuter}\ and\ \citenamefont
  {Saueressig}(2019)}]{reuter_saueressig_2019}%
  \BibitemOpen
  \bibfield  {author} {\bibinfo {author} {\bibfnamefont {M.}~\bibnamefont
  {Reuter}}\ and\ \bibinfo {author} {\bibfnamefont {F.}~\bibnamefont
  {Saueressig}},\ }\href {https://doi.org/10.1017/9781316227596} {\emph
  {\bibinfo {title} {Quantum Gravity and the Functional Renormalization Group:
  The Road towards Asymptotic Safety}}},\ Cambridge Monographs on Mathematical
  Physics\ (\bibinfo  {publisher} {Cambridge University Press},\ \bibinfo
  {year} {2019})\BibitemShut {NoStop}%
\bibitem [{\citenamefont {Jacobson}\ and\ \citenamefont
  {Satz}(2013)}]{jacobson_satz}%
  \BibitemOpen
  \bibfield  {author} {\bibinfo {author} {\bibfnamefont {T.}~\bibnamefont
  {Jacobson}}\ and\ \bibinfo {author} {\bibfnamefont {A.}~\bibnamefont
  {Satz}},\ }\bibfield  {title} {\bibinfo {title} {Black hole entanglement
  entropy and the renormalization group},\ }\href
  {https://doi.org/10.1103/PhysRevD.87.084047} {\bibfield  {journal} {\bibinfo
  {journal} {Phys. Rev. D}\ }\textbf {\bibinfo {volume} {87}},\ \bibinfo
  {pages} {084047} (\bibinfo {year} {2013})},\ \Eprint
  {https://arxiv.org/abs/hep-th/1212.6824} {arXiv:hep-th/1212.6824}
  \BibitemShut {NoStop}%
\bibitem [{\citenamefont {Wald}(1994)}]{wald1994quantum}%
  \BibitemOpen
  \bibfield  {author} {\bibinfo {author} {\bibfnamefont {R.~M.}\ \bibnamefont
  {Wald}},\ }\href@noop {} {\emph {\bibinfo {title} {Quantum Field Theory in
  Curved Spacetime and Black Hole Thermodynamics}}}\ (\bibinfo  {publisher}
  {University of Chicago Press},\ \bibinfo {year} {1994})\BibitemShut {NoStop}%
\bibitem [{\citenamefont {Calabrese}\ and\ \citenamefont
  {Cardy}(2004)}]{Calabrese_2004}%
  \BibitemOpen
  \bibfield  {author} {\bibinfo {author} {\bibfnamefont {P.}~\bibnamefont
  {Calabrese}}\ and\ \bibinfo {author} {\bibfnamefont {J.~L.}\ \bibnamefont
  {Cardy}},\ }\bibfield  {title} {\bibinfo {title} {{Entanglement entropy and
  quantum field theory}},\ }\href
  {https://doi.org/10.1088/1742-5468/2004/06/P06002} {\bibfield  {journal}
  {\bibinfo  {journal} {J. Stat. Mech.}\ }\textbf {\bibinfo {volume} {0406}},\
  \bibinfo {pages} {P06002} (\bibinfo {year} {2004})},\ \Eprint
  {https://arxiv.org/abs/hep-th/0405152} {arXiv:hep-th/0405152} \BibitemShut
  {NoStop}%
\bibitem [{\citenamefont {Casini}\ and\ \citenamefont
  {Huerta}(2009)}]{Casini_2009}%
  \BibitemOpen
  \bibfield  {author} {\bibinfo {author} {\bibfnamefont {H.}~\bibnamefont
  {Casini}}\ and\ \bibinfo {author} {\bibfnamefont {M.}~\bibnamefont
  {Huerta}},\ }\bibfield  {title} {\bibinfo {title} {{Entanglement entropy in
  free quantum field theory}},\ }\href
  {https://doi.org/10.1088/1751-8113/42/50/504007} {\bibfield  {journal}
  {\bibinfo  {journal} {J. Phys. A}\ }\textbf {\bibinfo {volume} {42}},\
  \bibinfo {pages} {504007} (\bibinfo {year} {2009})},\ \Eprint
  {https://arxiv.org/abs/hep-th/0905.2562} {arXiv:hep-th/0905.2562}
  \BibitemShut {NoStop}%
\bibitem [{\citenamefont {R\'enyi}(1961)}]{renyi1961}%
  \BibitemOpen
  \bibfield  {author} {\bibinfo {author} {\bibfnamefont {A.}~\bibnamefont
  {R\'enyi}},\ }\bibfield  {title} {\bibinfo {title} {On measures of entropy
  and information},\ }in\ \href
  {https://projecteuclid.org/euclid.bsmsp/1200512181} {\emph {\bibinfo
  {booktitle} {Proceedings of the Fourth Berkeley Symposium on Mathematical
  Statistics and Probability, Volume 1: Contributions to the Theory of
  Statistics}}}\ (\bibinfo  {publisher} {University of California Press},\
  \bibinfo {address} {Berkeley, Calif.},\ \bibinfo {year} {1961})\ pp.\
  \bibinfo {pages} {547--561}\BibitemShut {NoStop}%
\bibitem [{\citenamefont {R\'enyi}(1965)}]{renyi1965}%
  \BibitemOpen
  \bibfield  {author} {\bibinfo {author} {\bibfnamefont {A.}~\bibnamefont
  {R\'enyi}},\ }\bibfield  {title} {\bibinfo {title} {On the foundations of
  information theory},\ }\href {http://www.jstor.org/stable/1401301} {\bibfield
   {journal} {\bibinfo  {journal} {Revue de l'Institut International de
  Statistique / Review of the International Statistical Institute}\ }\textbf
  {\bibinfo {volume} {33}},\ \bibinfo {pages} {1} (\bibinfo {year}
  {1965})}\BibitemShut {NoStop}%
\bibitem [{\citenamefont {Nesterov}\ and\ \citenamefont
  {Solodukhin}(2011)}]{Nesterov_solodukhin}%
  \BibitemOpen
  \bibfield  {author} {\bibinfo {author} {\bibfnamefont {D.}~\bibnamefont
  {Nesterov}}\ and\ \bibinfo {author} {\bibfnamefont {S.~N.}\ \bibnamefont
  {Solodukhin}},\ }\bibfield  {title} {\bibinfo {title} {{Gravitational
  effective action and entanglement entropy in UV modified theories with and
  without Lorentz symmetry}},\ }\href
  {https://doi.org/10.1016/j.nuclphysb.2010.08.006} {\bibfield  {journal}
  {\bibinfo  {journal} {Nucl. Phys. B}\ }\textbf {\bibinfo {volume} {842}},\
  \bibinfo {pages} {141} (\bibinfo {year} {2011})},\ \Eprint
  {https://arxiv.org/abs/hep-th/1007.1246} {arXiv:hep-th/1007.1246}
  \BibitemShut {NoStop}%
\bibitem [{\citenamefont {Fursaev}\ and\ \citenamefont
  {Solodukhin}(1995)}]{Fursaev_1995}%
  \BibitemOpen
  \bibfield  {author} {\bibinfo {author} {\bibfnamefont {D.~V.}\ \bibnamefont
  {Fursaev}}\ and\ \bibinfo {author} {\bibfnamefont {S.~N.}\ \bibnamefont
  {Solodukhin}},\ }\bibfield  {title} {\bibinfo {title} {{On the description of
  the Riemannian geometry in the presence of conical defects}},\ }\href
  {https://doi.org/10.1103/PhysRevD.52.2133} {\bibfield  {journal} {\bibinfo
  {journal} {Phys. Rev. D}\ }\textbf {\bibinfo {volume} {52}},\ \bibinfo
  {pages} {2133} (\bibinfo {year} {1995})},\ \Eprint
  {https://arxiv.org/abs/hep-th/9501127} {arXiv:hep-th/9501127} \BibitemShut
  {NoStop}%
\bibitem [{\citenamefont {Iyer}\ and\ \citenamefont {Wald}(1995)}]{IW95}%
  \BibitemOpen
  \bibfield  {author} {\bibinfo {author} {\bibfnamefont {V.}~\bibnamefont
  {Iyer}}\ and\ \bibinfo {author} {\bibfnamefont {R.~M.}\ \bibnamefont
  {Wald}},\ }\bibfield  {title} {\bibinfo {title} {A comparison of {Noether}
  charge and {Euclidean} methods for computing the entropy of stationary black
  holes},\ }\href {https://doi.org/10.1103/PhysRevD.52.4430} {\bibfield
  {journal} {\bibinfo  {journal} {Phys. Rev. D}\ }\textbf {\bibinfo {volume}
  {52}},\ \bibinfo {pages} {4430} (\bibinfo {year} {1995})},\ \Eprint
  {https://arxiv.org/abs/gr-qc/9503052} {arXiv:gr-qc/9503052} \BibitemShut
  {NoStop}%
\bibitem [{Note1()}]{Note1}%
  \BibitemOpen
  \bibinfo {note} {It interesting to note that Eq.~(\ref {generalWald}) is
  derived on-shell while the conical method is an off-shell procedure which is
  valid in any spacetime with a bifurcate Killing horizon. For the relation
  between the on- and off-shell approaches, see for instance, Ref.~\cite
  {Solodukhin_review}}\BibitemShut {NoStop}%
\bibitem [{Note2()}]{Note2}%
  \BibitemOpen
  \bibinfo {note} {It is important to note that restricting to such a
  submanifold in the theory space will induce a restriction of the entropy flow
  to a submanifold in the space of entropies.}\BibitemShut {Stop}%
\bibitem [{\citenamefont {Hartle}\ and\ \citenamefont
  {Hawking}(1976)}]{harte_hawking_original}%
  \BibitemOpen
  \bibfield  {author} {\bibinfo {author} {\bibfnamefont {J.~B.}\ \bibnamefont
  {Hartle}}\ and\ \bibinfo {author} {\bibfnamefont {S.~W.}\ \bibnamefont
  {Hawking}},\ }\bibfield  {title} {\bibinfo {title} {Path-integral derivation
  of black-hole radiance},\ }\href {https://doi.org/10.1103/PhysRevD.13.2188}
  {\bibfield  {journal} {\bibinfo  {journal} {Phys. Rev. D}\ }\textbf {\bibinfo
  {volume} {13}},\ \bibinfo {pages} {2188} (\bibinfo {year}
  {1976})}\BibitemShut {NoStop}%
\bibitem [{\citenamefont
  {Jacobson}(1994{\natexlab{a}})}]{jacobson_hartle_hawking_vacua}%
  \BibitemOpen
  \bibfield  {author} {\bibinfo {author} {\bibfnamefont {T.}~\bibnamefont
  {Jacobson}},\ }\bibfield  {title} {\bibinfo {title} {Note on hartle-hawking
  vacua},\ }\href {https://doi.org/10.1103/PhysRevD.50.R6031} {\bibfield
  {journal} {\bibinfo  {journal} {Phys. Rev. D}\ }\textbf {\bibinfo {volume}
  {50}},\ \bibinfo {pages} {R6031} (\bibinfo {year} {1994}{\natexlab{a}})},\
  \Eprint {https://arxiv.org/abs/gr-qc/9407022} {arXiv:gr-qc/9407022}
  \BibitemShut {NoStop}%
\bibitem [{\citenamefont {Kay}\ and\ \citenamefont {Wald}(1991)}]{WK91}%
  \BibitemOpen
  \bibfield  {author} {\bibinfo {author} {\bibfnamefont {B.~S.}\ \bibnamefont
  {Kay}}\ and\ \bibinfo {author} {\bibfnamefont {R.~M.}\ \bibnamefont {Wald}},\
  }\bibfield  {title} {\bibinfo {title} {Theorems on the uniqueness and thermal
  properties of stationary, nonsingular, quasifree states on spacetimes with a
  bifurcate killing horizon},\ }\href
  {https://doi.org/https://doi.org/10.1016/0370-1573(91)90015-E} {\bibfield
  {journal} {\bibinfo  {journal} {Physics Reports}\ }\textbf {\bibinfo {volume}
  {207}},\ \bibinfo {pages} {49 } (\bibinfo {year} {1991})}\BibitemShut
  {NoStop}%
\bibitem [{\citenamefont {Percacci}(2017)}]{Percacci_Book}%
  \BibitemOpen
  \bibfield  {author} {\bibinfo {author} {\bibfnamefont {R.}~\bibnamefont
  {Percacci}},\ }\href {https://doi.org/10.1142/10369} {\emph {\bibinfo {title}
  {{An Introduction to Covariant Quantum Gravity and Asymptotic Safety}}}},\
  \bibinfo {series} {100 Years of General Relativity}, Vol.~\bibinfo {volume}
  {3}\ (\bibinfo  {publisher} {World Scientific},\ \bibinfo {year}
  {2017})\BibitemShut {NoStop}%
\bibitem [{Note3()}]{Note3}%
  \BibitemOpen
  \bibinfo {note} {Even if we had taken a truncation subspace that included
  such terms, our conclusions would be left unchanged. In this case, the
  quantum corrections would only renormalize the $c_j$, $j\in \protect \mathbb
  {N}$, and would not mix the entanglement entropy with other entropy terms.
  Hence, the inclusion of such terms would only render the expressions more
  lengthy and would not add anything relevant to the discussion. For this
  reason, we neglect them without any conceptual loss.}\BibitemShut {Stop}%
\bibitem [{Note4()}]{Note4}%
  \BibitemOpen
  \bibinfo {note} {If $p^{2}$ are the eigenvalues of $\nabla ^{2}$, as
  $R_k(p^2)$ vanishes for $p^2>k^2,$ it leaves the contribution of the
  high-energy modes to ${\protect \rm sTr}_{\protect \mathcal {M}_\alpha
  }\protect \qopname \relax o{log}\left (P_k\left (\nabla ^{2}\right
  )+m^2_{\Lambda }\right )$ untouched while it suppresses the contribution of
  the low-energy modes, $p^2<k^2$. Hence, Eq.~(\ref {W_alpha_k}) gives the
  (low-energy) effective action at a scale $k$ (it is easy to see that $W_k$
  indeed vanishes for $p^2>k^2$). A similar reasoning is valid for the other
  terms we analyze next.}\BibitemShut {Stop}%
\bibitem [{\citenamefont {{Sakharov}}(1968)}]{S68}%
  \BibitemOpen
  \bibfield  {author} {\bibinfo {author} {\bibfnamefont {A.~D.}\ \bibnamefont
  {{Sakharov}}},\ }\bibfield  {title} {\bibinfo {title} {{Vacuum Quantum
  Fluctuations in Curved Space and the Theory of Gravitation}},\ }\href@noop {}
  {\bibfield  {journal} {\bibinfo  {journal} {Soviet Physics Doklady}\ }\textbf
  {\bibinfo {volume} {12}},\ \bibinfo {pages} {1040} (\bibinfo {year}
  {1968})}\BibitemShut {NoStop}%
\bibitem [{\citenamefont {Jacobson}(1994{\natexlab{b}})}]{J94a}%
  \BibitemOpen
  \bibfield  {author} {\bibinfo {author} {\bibfnamefont {T.}~\bibnamefont
  {Jacobson}},\ }\href@noop {} {\bibinfo {title} {Black hole entropy and
  induced gravity}} (\bibinfo {year} {1994}{\natexlab{b}}),\ \Eprint
  {https://arxiv.org/abs/gr-qc/9404039} {arXiv:gr-qc/9404039} \BibitemShut
  {NoStop}%
\bibitem [{\citenamefont {Jacobson}(1995)}]{J95}%
  \BibitemOpen
  \bibfield  {author} {\bibinfo {author} {\bibfnamefont {T.}~\bibnamefont
  {Jacobson}},\ }\bibfield  {title} {\bibinfo {title} {Thermodynamics of
  {Spacetime}: {The} {Einstein} {Equation} of {State}},\ }\href
  {https://doi.org/10.1103/PhysRevLett.75.1260} {\bibfield  {journal} {\bibinfo
   {journal} {Phys. Rev. Lett.}\ }\textbf {\bibinfo {volume} {75}},\ \bibinfo
  {pages} {1260} (\bibinfo {year} {1995})},\ \Eprint
  {https://arxiv.org/abs/gr-qc/9504004} {arXiv:gr-qc/9504004} \BibitemShut
  {NoStop}%
\end{thebibliography}%
\end{document}